\documentclass[letterpaper,twocolumn,10pt]{article}
\usepackage{usenix}
\pdfoutput=1 

\usepackage{hyperref}
\usepackage{enumitem}
\usepackage{verbatim}
\usepackage[nodisplayskipstretch]{setspace}
\usepackage{subcaption} % for subfigures
\usepackage{algorithm}
\usepackage[noend]{algpseudocode}
\usepackage{color}

\newcommand{\ms}[1]{{{\color{red}[ #1 -- Marco]}}}
\newcommand{\nn}[1]{{{\color{red}[ #1 -- Nathan]}}}

\usepackage{amsmath}
\usepackage{amsthm}
\usepackage{pifont}% http://ctan.org/pkg/pifont
\usepackage{subcaption}
\usepackage{mwe}
\usepackage[title]{appendix}
\newcommand{\spara}[1]{\noindent \textbf{#1. }}
\newcommand{\sparalight}[1]{\noindent \emph{#1 -- }}
\newtheorem{theorem}{Theorem}[section]
\newtheorem{lemma}[theorem]{Lemma}

\newtheorem{definition}[theorem]{Definition}
\hypersetup{
    colorlinks=true,
    linkcolor=blue,
    filecolor=magenta,      
    urlcolor=cyan,
    pdftitle={Overleaf Example},
    pdfpagemode=FullScreen,
    }

\begin{document}

%%
%% The "title" command has an optional parameter,
%% allowing the author to define a "short title" to be used in page headers.
%\title{Tuning the Tail Latency of Distributed Queries Using Replication}
\title{Tuning the Tail Latency of Distributed Queries Using Replication}

%\author{}
\author{Nathan Ng, Hung Le, Marco Serafini\\ University of Massachusetts Amherst}

\maketitle

%%
%% The abstract is a short summary of the work to be presented in the
%% article.
\begin{abstract}
  Querying graph data with low latency is important in application domains such as social networks and knowledge graphs.
  Graph queries perform multiple hops between vertices.
  When data is partitioned and stored across multiple servers, queries executing at one server often need to hop to vertices stored by another server. 
  Such \emph{distributed traversals} represent a performance bottleneck for low-latency queries.
  To reduce query latency, one can replicate remote data to make distributed traversals unnecessary, but replication is expensive and should be minimized.
  
  In this paper, we introduce the problem of finding data replication schemes that satisfy arbitrary user-defined query latency constraints with minimal replication cost.
  We propose a novel workload model to express data access causality, propose a family of heuristics, and introduce non-trivial sufficient conditions for their correctness.
  Our evaluation using two representative benchmarks shows that our algorithms enable fine-tuning query latency with data replication and can find sweet spots in the latency/replication design space.
\end{abstract}

\section{Introduction}
\label{sec:intro}
In many important domains, such as social media, search engines, and question answering (Q\&A) systems, it is common to model structured data as a graph.
Applications in these domains often operate on large graphs stored on a distributed system and generate workloads with end-to-end tail latency constraints in the order of milliseconds~\cite{A1,tao,unicorn,Wukong,Pragh,social_hash,linkbench,trinity_rdf,chiller,managing_large_dynamic_graphs_efficiently,DistDGL}.
For example, Facebook uses a graph to represent critical data such as users, posts, or comments, and queries the graph to render web pages and to serve user searches~\cite{tao,linkbench,unicorn}.
The Bing search engine queries its knowledge graph to render pages~\cite{A1}.
Q\&A systems powering Amazon's and Alibaba's personal assistants query their knowledge graphs~\cite{ubiquity}.
To support low-latency graph queries, several distributed in-memory graph data management systems have been proposed and used in production~\cite{trinity, tao, trinity_rdf, A1, Wukong, Pragh, gtran, grasper, neo4j_linkedin}.
These systems speed up general query execution, but supporting tail latency bounds still remains an open research problem.

\begin{figure}[t!]
  \centering
  \includegraphics[width=.6 \columnwidth]{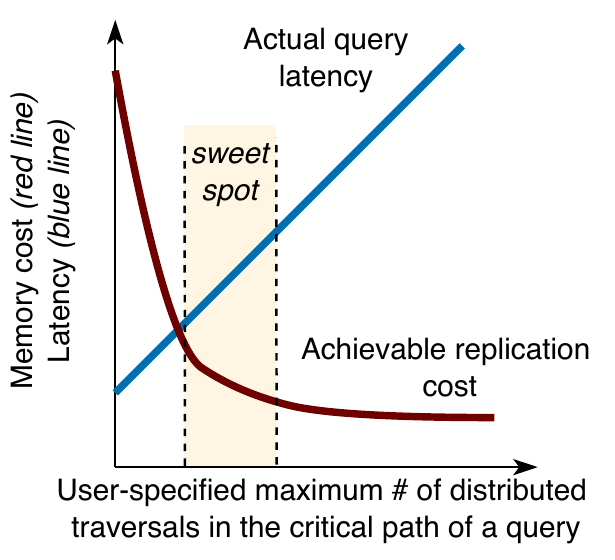}
  \caption{Tradeoff of fine-tuning latency using replication.}
  \label{fig:tradeoff}
\end{figure}

In this paper, we address for the first time the problem of \emph{ensuring that distributed queries respect an arbitrary user-defined bound on tail latency}.
This is a stricter requirement than existing work that reduces latency across an entire workload using workload-aware data placement and migration techniques but does not ensure tail  latency bounds
\cite{schism,social_hash,serafini2014accordion,taft2014store,Clay,Pragh,hermes,squall,morphus,
sword,JECB,graph_partitioning,managing_large_dynamic_graphs_efficiently,accelerate_SPARQL}.
In this first work addressing this  problem, we focus on \emph{low-latency read} queries, which dominate the graph workloads we discussed and often have the strictest low-latency constraints.
When the graph is partitioned and stored across distributed servers, ensuring that these queries reliably respect their tail latency constraint is challenging.
This is because a query executing at a server may require \emph{distributed traversals}, that is, traverse edges to destination vertices stored by other servers.
Low-latency read queries perform very few data accesses and minimal computation, so their latency is determined by their \emph{data access locality}.
Remote data accesses are much slower than local ones due to the fundamental physics of data locality. 
Prior work showed that even with graph databases using advanced networking technologies such as Remote Direct Memory Access (RDMA), local accesses are 20x-100x faster than remote accesses~\cite{A1,Pragh,Wukong,gtran,grasper} and executing 2-hop graph queries locally to a single server is 30X faster than when data is distributed over an 8-node RDMA-based graph database~\cite{Pragh}.

Our evaluation echoes these findings and shows that the latency of a low-latency distributed read query is a function of the maximum number of distributed traversals on its critical path.
To bound the latency of these queries, one can eliminate distributed traversals by substituting remote data accesses with local accesses to \emph{replicas} of the required data. 
However, replication is memory-expensive and must be carefully applied. 
Therefore, understanding the tradeoff between latency and replication cost is crucial.

This paper proposes the first algorithms that enable users to navigate the tradeoff between the benefit of setting stricter tail latency bounds and the replication cost of enforcing those bounds (see Figure~\ref{fig:tradeoff}).
At one end of this tradeoff, a user could require that all queries be executed locally at a single server, without distributed traversals.
This results in the lowest latency but entails a very high replication cost even when using an oracle with perfect knowledge of the workload, as we show.
Previous work has aimed to reduce the cost of single-site execution for certain workloads~\cite{SPAR,MorphoSys}.
Our work takes a different direction: it shows for the first time that relaxing the latency constraint and allowing for a very small number of distributed traversals can reduce memory cost significantly.
%We observed in our evaluation that the cost drops steeply even when relaxing the latency constraint only slightly (see the red line in Figure~\ref{fig:tradeoff}).
We observed that by fine-tuning the latency bound, we could find a sweet spot between performance and replication cost. 

Our first contribution is formalizing the new \emph{latency-bound replication problem}.
Our goal is to generate a replication scheme that respects an arbitrary used-defined latency bound while minimizing replication cost. 
The latency of a query is determined by its slowest chain of \emph{causally-dependent data accesses}, which must be performed sequentially.
For example, graph hops introduce causal dependencies: accessing the neighbors of a vertex is only possible after accessing the adjacency list of the vertex.
%Constraining latency requires reasoning in terms of the worst-case access paths of a query.
We introduce the notion of \emph{causal access paths} to represent the causal dependencies between data accesses, which are not captured by the workload graph models used by prior work on distributed data placement~\cite{schism,social_hash,Clay}.
%\red{The simple tree structure of the causal access trees makes it easier to design a replication scheme by focusing on each root-to-leaf path of the trees in isolation.}
Our formalism is general and applies to a wide range of query execution systems, independent of how they express queries (e.g., using a relational or graph query language), how they perform distributed query optimization, and how they store and shard data.

%\nn{represented or stored? Should be structured (relational) or unstructured (graph)?}.
%The system model pushes subqueries to remote servers to maximize query locality and replication flexibility, rather than fetching data. 
%We also model and implement graph query executors that maximize query locality and leverage data replication by pushing subqueries to remote servers, rather than fetching data.
%Navigating the space of solutions to this problem introduces a number of novel and interesting technical challenges.

%Any data placement and replication scheme must be coupled with a routing algorithm to locate data during online query execution.
%Routing must be efficient and requires only compact routing data structures.
%A data replication scheme that requires different queries to access different replicas of the same object to meet their latency bound might generate huge routing tables.
%We propose a \emph{thrifty routing} scheme that does not require storing any additional routing information.
%We then propose data replication schemes that are tailored to thrifty routing and still enforce the mandated latency bounds.

\begin{figure*}[t!]
  \centering
  \begin{subfigure}[b]{0.21\textwidth}
    \centering
    \includegraphics[width=\textwidth]{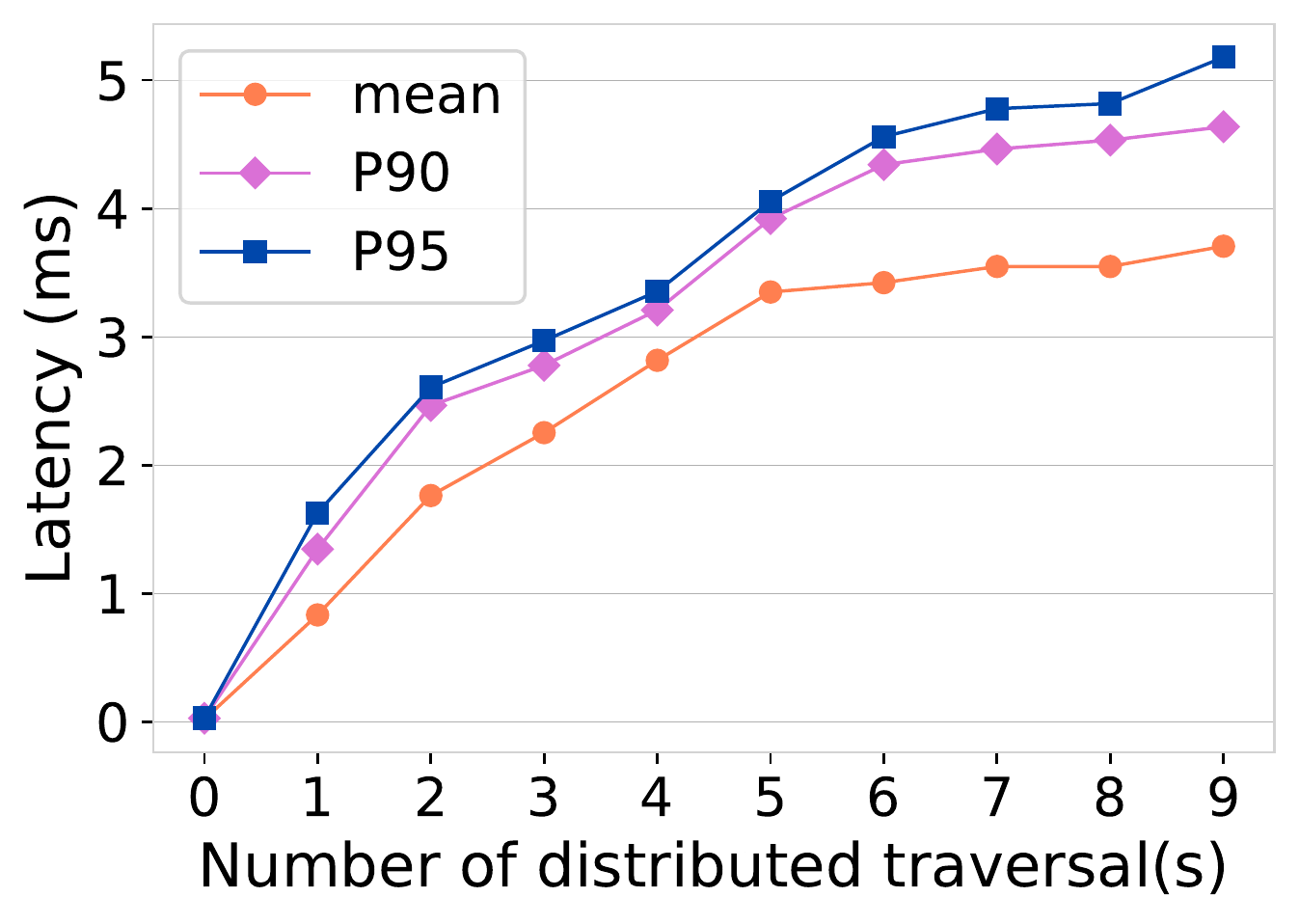}
    \caption{Query latency vs. number of  distributed traversals in critical path}
    \label{fig:hop_vs_latency}
  \end{subfigure}
  \hfill
  \begin{subfigure}[b]{0.21\textwidth}
      \centering
      \includegraphics[width=\textwidth]{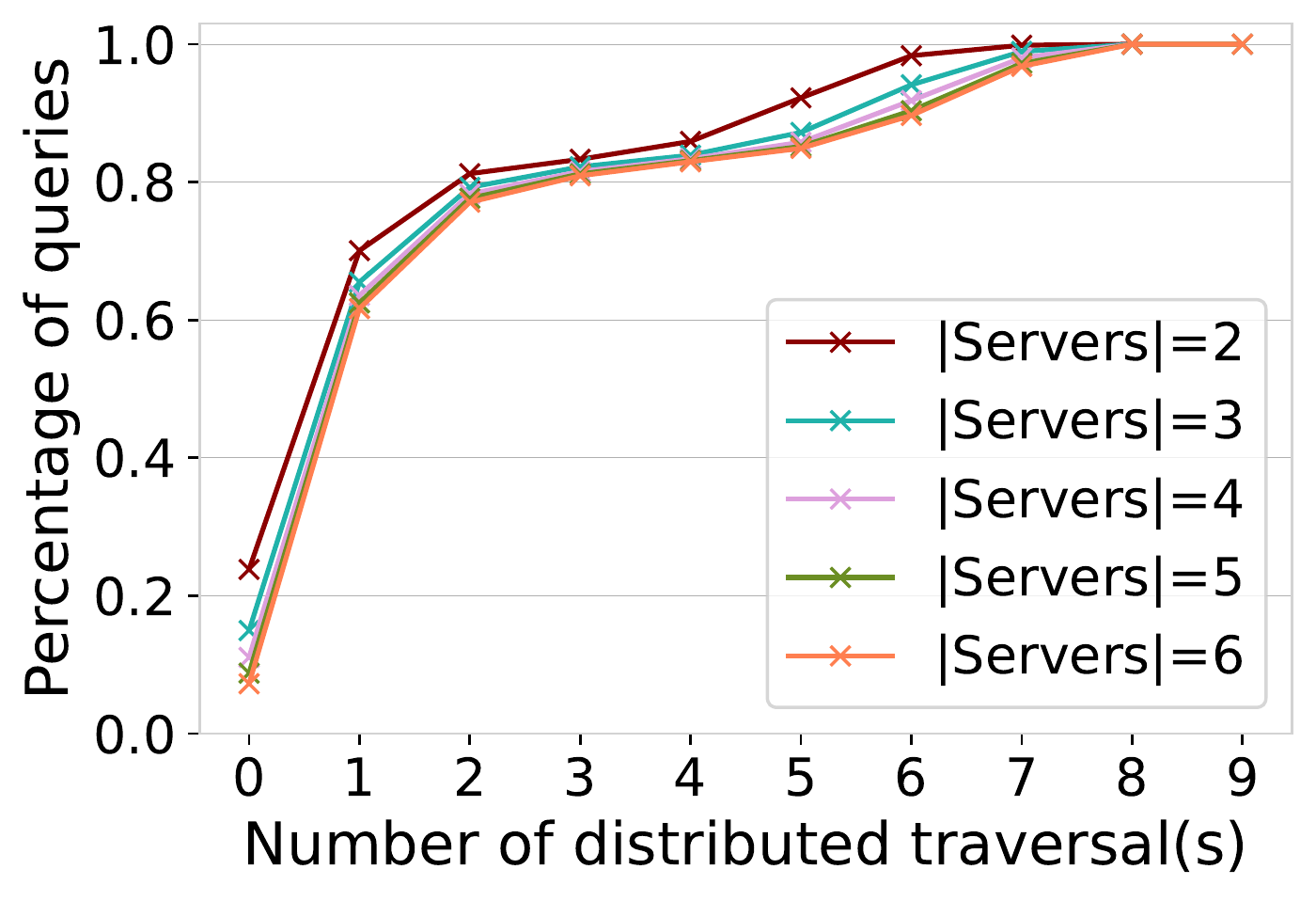}
      \caption{CDF of distributed traversals required per query with hash sharding.}
      \label{fig:remote_per_q_hash}
  \end{subfigure}
  \hfill
  \begin{subfigure}[b]{0.21\textwidth}
      \centering
      \includegraphics[width=\textwidth]{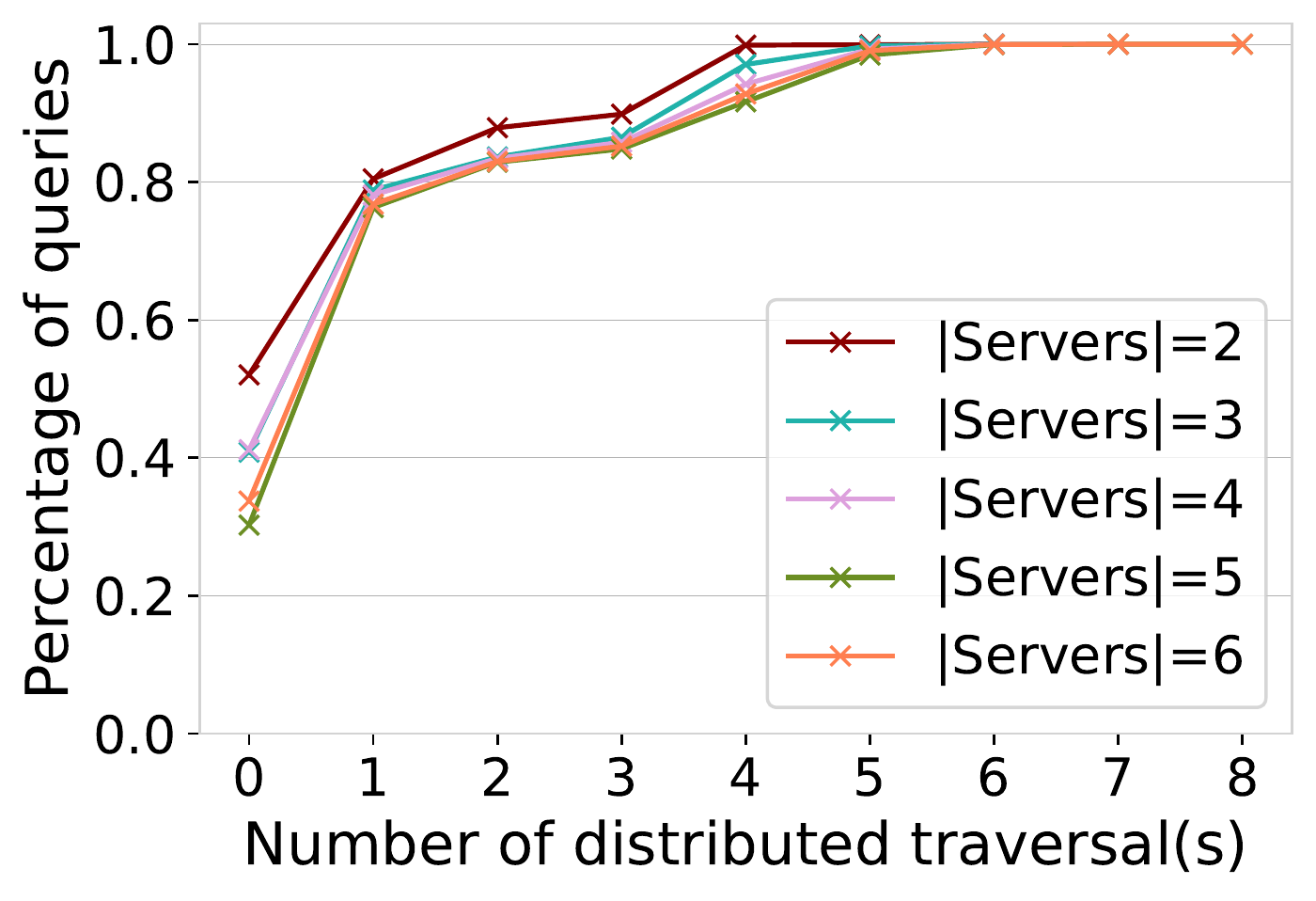}
      \caption{CDF of distributed traversals required per query with min-cut sharding.}
      \label{fig:remote_per_q_metis}
  \end{subfigure}
  \hfill
  \begin{subfigure}[b]{0.21\textwidth}
    \centering
    \includegraphics[width=\textwidth]{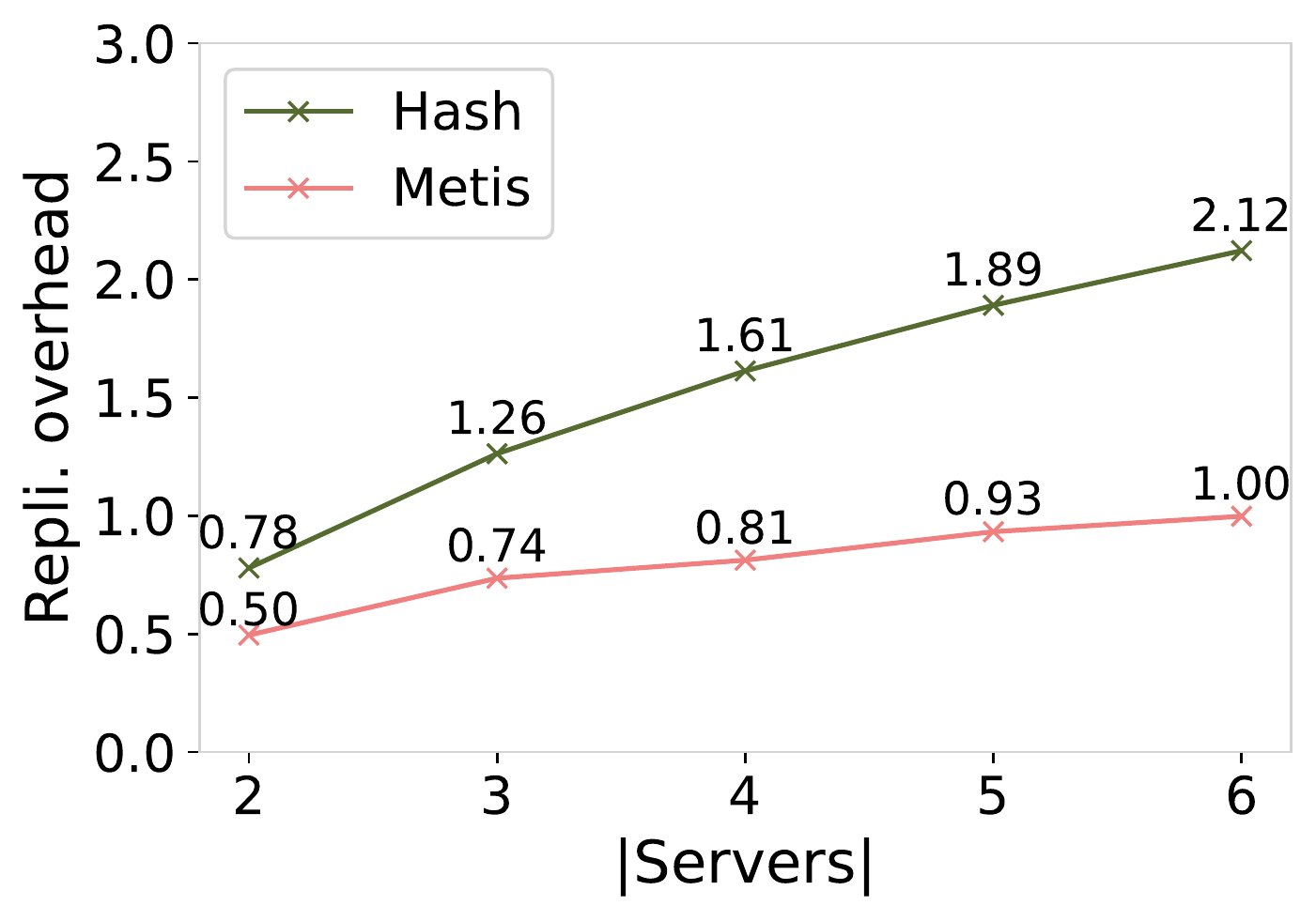}
    \caption{Replication overhead of single-site execution with different data placements.}
    \label{fig:k0_replication_cost}
  \end{subfigure}
  \label{fig:short_queries_remote_per_q}
  \vspace{-0.5\baselineskip}
  \caption{Distributed multi-hop queries in the LDBC SNB interactive workload.}
  \vspace{-0.5em}
\end{figure*}

We show that the latency-bound replication problem is NP-hard.
Designing greedy algorithms is intuitive, but  proving their correctness is surprisingly non-trivial because, somewhat counter-intuitively, greedily adding new replicas may introduce violations of the latency constraint of some previously optimized queries.
We then introduce a property called \emph{latency-robustness} and show that, if enforced by the replication algorithm, it avoids such problems.
Finally, we discuss how to adapt the replication scheme when there are system reconfigurations or server faults.

Our evaluation shows that our algorithms enable fine-tuning latency (both average and tail) and trading latency guarantees with data replication cost.
We consider two diverse benchmarks: LDBC SNB, and specifically its low-latency user-facing read queries~\cite{ldbc}, and node-wise neighborhood sampling, which is a critical bottleneck for distributed mini-batch training of Graph Neural Networks (GNNs)~\cite{DistDGL}. 
%To show the generality of the approach, we also consider a different type of workload: distributed node-wise neighborhood sampling for Graph Neural Network training over graphs from the Open Graph Benchmark \cite{OGB}.
%We consider neighborhood sampling as the workload, which is performed online at each training iteration to obtain mini-batches \cite{DistDGL}. 
%Distributed GNN training systems like DistDGL perform distributed graph sampling online at each training iteration to obtain mini-batches \cite{DistDGL}.
%To reduce GNN training time, sampling must be efficient \cite{nextdoor, case}.
%We show that the replication criteria used in DistDGL can be seen as a special case of our framework and that our algorithms can achieve a lower replication cost compared to DistDGL.
Setting the latency constraints in these workloads can vary the tail and average query latency by a factor of up to $120\times$ and the overall storage cost by up to $3\times$.
It is therefore critical to fine-tune this parameter to find the right latency-replication tradeoff.
%Our selective data replication algorithms enable users to control both average and tail latency.
%We find that by relaxing the latency bound to allow some amount of traversals, we can significantly reduce the replication cost while still being able to control query latencies
Our replication algorithm allows relaxing the latency constraint to reduce replication cost until we reach a sweet spot in both workloads.
We can tune latencies in the range between sub-milliseconds and single-digit milliseconds, in line with the performance requirements of production applications such as knowledge graphs for search engines~\cite{A1}.
Our offline algorithm is fast enough to be practical as a one-off cost: it can analyze a workload and generate a replication scheme in seconds for million-scale graphs and in tens of minutes for billion-scale graphs.
% The algorithm is effective under a variety of sharding schemes and can be extended to incrementally update the replication scheme with a low replication cost. 
%In GNN sampling workloads, our algorithms can achieve the same latency guarantees as DistDGL's replication scheme with lower replication cost. \nn{remove this and not stress this part?}

In summary, our key contributions are:
\begin{enumerate}[parsep=1.5pt,topsep=1.5pt]
  \item We formalize for the first time the \emph{latency-bound replication problem} to navigate the latency-replication tradeoff, and introduce \emph{causal access paths} to model query latency. 
  \item We prove that the problem is NP-hard, propose the first greedy replication algorithm for the problem, and introduce a non-trivial sufficient condition for correctness called \emph{latency-robustness}.
  \item We extend our algorithm to update the replication scheme in reaction to system reconfigurations or faults.
  \item Using two diverse benchmarks, our evaluation shows that our algorithm effectively achieves the desired tail latency bounds, significantly reduces the replication cost when the user even slightly relaxes the latency constraints, has low running time, is effective under different sharding schemes, and can incrementally update the replication scheme with a moderate replication cost.
\end{enumerate}
This paper proposes a novel and general algorithmic framework for supporting tail latency bounds using replication.
While the focus of this first effort is on low-latency read graph queries, we believe that our work opens the way to several future extensions.

\section{Background and Motivation}
\label{sec:background}
To understand the behavior of read queries with the strictest latency constraints in graph workloads, we analyzed LDBC SNB's interactive short read query benchmark~\cite{ldbc} (see Section~\ref{sec:eval} for the details of the experimental setup). 
We found that the latency of a query is a function of the number of distributed traversals on its critical path.
We also found that enforcing single-site query execution results in a very high replication cost.

\spara{Query latency and distributed traversals}
We find that queries with strict low-latency constraints only access few data objects and perform minimal computation.
Remote data accesses are inherently slower than local ones due to the fundamental physics of data locality.
For generality, our evaluation focuses on common Gigabit networks, but our findings echo the results of existing work on RDMA-based graph databases, which are discussed in Section~\ref{sec:intro}.

Figure~\ref{fig:hop_vs_latency} shows that as the number of distributed traversals in the critical path of a query grows, the both tail and average latency grow almost linearly.
The number of distributed traversals required may be large, depending on the sharding scheme. 
Figures~\ref{fig:remote_per_q_hash} and ~\ref{fig:remote_per_q_metis} show the CDF of the number of distributed traversals required per query with varying numbers of servers in the system.
With distributed hash partitioning, which is commonly used by in-memory graph databases such as A1~\cite{A1} and Wukong~\cite{Wukong}, 30-40\% of queries require more than one traversal (see Figure~\ref{fig:remote_per_q_hash}). 
Note that the number of distributed traversals can be larger than the number of servers because a query may traverse between two servers multiple times.
Using an offline min-cut partitioning algorithm such as Metis~\cite{METIS} to partition the data graph while minimizing edge cuts across servers can reduce the number of distributed traversals, as shown in Figure~\ref{fig:remote_per_q_metis}. 
However, even with this more complex and costly data placement, there still exists a long tail of queries that require a large amount of distributed traversals.

\spara{The high cost of single-site query execution}
A simple approach to constrain latency is to replicate data until no distributed traversal is necessary for any query. 
A query is initially routed to some server and then executed locally only by that server.
% (a.k.a., providing a $0$-traversal guarantee) \cite{SPAR}. 
We evaluated the memory cost of this approach when using an oracle that exploits perfect knowledge of the workload to minimize replication.
First, we run the workload and record accessed data query by query.
Next, for each query, we replicate only the data actually accessed to ensure local execution. 
We report the replication overhead of using this perfect oracle in Figure \ref{fig:k0_replication_cost}. 
With Metis partitioning, the system has to store twice as much data as the original graph in a cluster of six servers. 
With hash partitioning, the memory requirement is more than three times the original one.

The SPAR~\cite{SPAR} and MorphoSys~\cite{MorphoSys} systems adopt the single-site query execution strategy and propose techniques to reduce the memory cost for certain workloads.
However, they cannot be applied to the workloads we consider in this paper.
SPAR assumes 1-hop graph workloads where queries visit a vertex and its immediate neighbors only.
It then shards and replicates data to ensure that the original copy of each vertex is co-located with the original or replica copy of each of its neighbors. 
Our work supports multi-hop read queries and more flexible latency constraints to reduce the memory cost, which is high in SPAR.
%It allows users to reduce replication cost by allowing some amount of distributed traversals while preserving user-defined tail latency bounds.

MorphoSys leverages data migration to enforce single-site execution for read and write queries. 
It requires that the read and write sets of each online query be known a priori, so the server executing the query knows immediately all the data the query will access and can fetch it, if needed, in at most one round of communication.
After that, the server executes the query locally.
Our approach targets workloads where MorphoSys' approach is not applicable: read queries which could require multiple causally-dependent distributed traversals since the read sets are not known a priori.

\spara{Goals of this work}
These results indicate that bounding the number of distributed traversals of a query using replication can control tail and average latency.
They also show that there is a need for reducing the high replication cost of ensuring single-site execution.
The goal of our work is to allow the user to slightly relax the strict requirement of single-site execution to drastically reduce the replication cost while still enforcing query latency bounds.

\section{Overview}
\label{sec:overview}
\subsection{Targeted Workloads and Systems}
\label{sec:models}

We now describe the workload and the query execution system models we use as inputs to our problem definition.

%This work considers multi-hop graph workloads with strict low-latency constraints.
%We focus on read-dominated workloads, which are common in many applications such as social networks and knowledge bases, among others \ms{refs}.
%Because of their strict latency constraints, these queries perform relatively few data accesses and minimal computation and are bottlenecked by distributed traversals, as we have shown previously.
%Updates must be ingested quickly but it is acceptable if reads observe slightly stale data.
%\ms{A1: (section 5) graph generated once a day}
%We now describe how our algorithms model a workload and a distributed query execution system.
%These models are fed as input to our replication algorithms.

\spara{Workload model}
We model a \emph{dataset} as a set of abstract \emph{objects}.
This paper focuses on graph databases, and we primarily consider a vertex and its adjacency list as an object.
However, the model is generic and applies to other types of databases.

\begin{figure}[t!]
    \begin{subfigure}[b]{.4\linewidth}
        \centering
        \includegraphics[width=.65\textwidth]{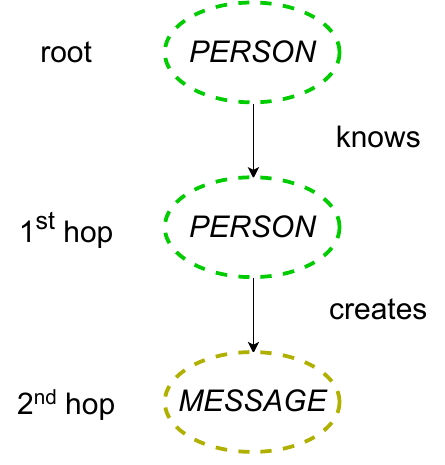}
        \caption{Multi-hop query type (as a labeled graph).}
        \label{fig:multihop_example}
    \end{subfigure}
    \hspace{5mm}
    \begin{subfigure}[b]{.4\linewidth}
        \centering
        \includegraphics[width=\textwidth]{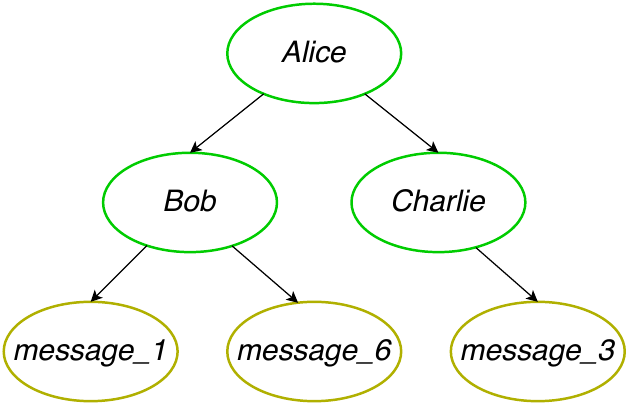}
        \caption{Causal access paths of a query instance.}
        \label{fig:causal_access_tree}
    \end{subfigure}
    \vspace{-0.5\baselineskip}
    \caption{Workload model: A query is modeled as a set of causal access paths.}
    \label{fig:example_query}
    \vspace{-1em}
\end{figure}

The \emph{workload} model describes which objects are accessed by each query and whether these accesses are causally-related.
We introduce the notion of \emph{causal access paths} to model queries.
Each query is modeled as one or more causal access paths.
Consider for example an instance of the query type of Figure~\ref{fig:example_query}.
Suppose the query starts by accessing Alice's vertex, which becomes the root of the causal access path.
Only after accessing the friends list of Alice can the query access the vertices associated with Bob and Charlie. 
In this case, the access to Alice's vertex happens before the accesses to Bob and Charile's vertices.
The causal dependency information captured by the causal access path is key to model the latency cost of a query.
Causally-dependent accesses must be sequential, whereas non-causally-dependent ones can be parallel.
The latency of a query is defined by the latency of its slowest root-to-leaf path of causally-related data accesses.
In our target workloads, the latency of a critical path is equal to the number of its distributed traversals, as we have shown in Section~\ref{sec:background}.

This workload model is general and applies to different query execution strategies and query languages.
As we will discuss, our greedy algorithms operate path-by-path and do not need to materialize the entire workload model.
%We will also discuss examples of workload analyzers to generate workload models in Section~\ref{sec:eval}.

Our modeling approach departs from prior work using \emph{workload graphs}, where data objects are represented by vertices that are connected by edges, or hyperedges, if they are accessed by the same query~\cite{schism,sword,JECB,graph_partitioning,Clay}.
To minimize the \emph{total number} of distributed traversals in a workload, one can partition a workload graph to minimize edge cut and then assign each partition to a different server.
However, to model the number of distributed traversals \emph{per-query},  it is key to capture the causal dependencies between data accesses in each query.
This information is not available in workload graphs but it is captured by causal access paths.

\begin{figure}[t!]
    \centering
    \includegraphics[width=.9\columnwidth]{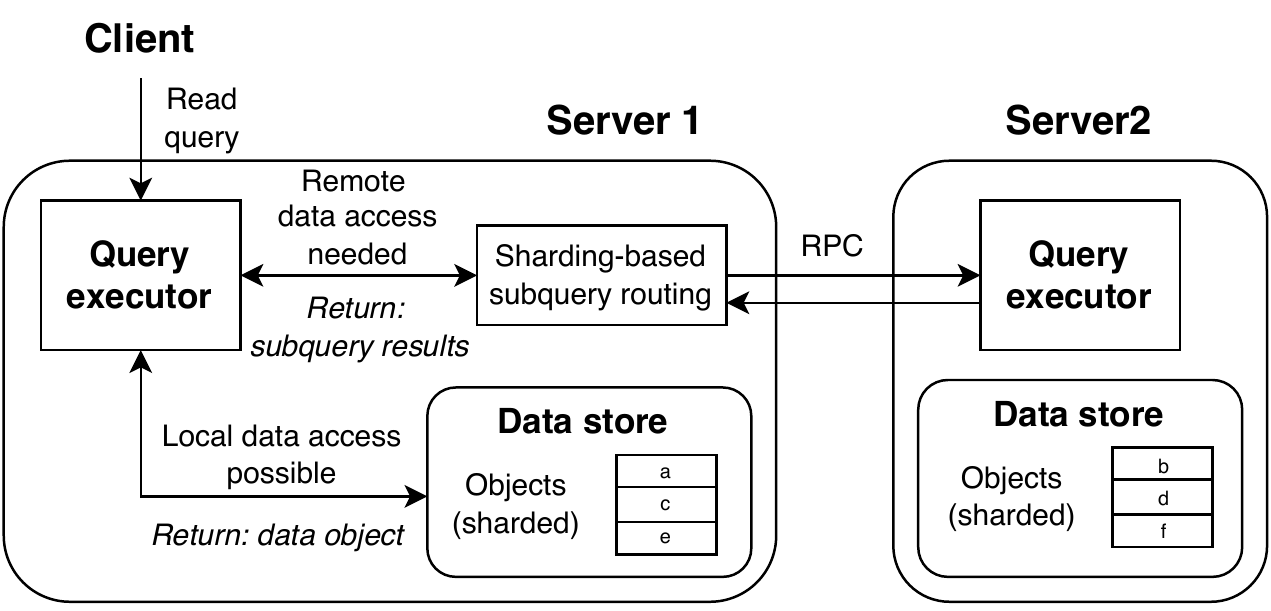}
    \caption{System model: Query execution system.}
    \label{fig:overview-diagram}
\end{figure}

\spara{System model}
We consider a distributed query execution system as depicted in Figure~\ref{fig:overview-diagram}.
We assume that the servers of the system have two main components: a \emph{data store} and a \emph{query executor}.
Before we run our algorithms, the in-memory store instance at each server contains one partition (or shard) of the dataset.
Many graph databases use hash partitioning to map vertices (i.e., objects) to servers. 
Some support using graph partitioning algorithms to store vertices in densely connected subgraphs on the same server.
Others analyze the workload to shard the data, as done by Social Hash~\cite{social_hash}.
We do not make assumptions about the \emph{sharding function} used by the system and treat it as an input to our problem. 
This generality makes our approach easier to adopt in existing systems; we illustrate this in our evaluation, where we stack our replication schemes on top of all the aforementioned sharding functions.

The second main component in a server is the query executor, which serves client queries by accessing the local in-memory store and cooperating with other servers' query executors.
Distributed execution of read queries uses nested Remote Procedure Calls (RPCs).
For example, in the query of Figure~\ref{fig:overview-diagram}, if a query executor at a server (say $s_1$) accesses Alice locally but has no local copy of Charlie, it sends an RPC to the server hosting Charlie's vertex (say $s_2$).
This is called \emph{sharding-based routing} in Figure~\ref{fig:overview-diagram}.
Server $s_2$ then continues the execution of the subquery from there: it finds the edges to the messages created by Charlie and accesses them.
If any of these messages is stored on a different server, $s_2$ invokes a nested RPC at that server.
At the end of the query execution, the coordinator gathers all the accessed data, computes aggregations if required, and returns the result to the client.
This subquery shipping approach avoids repeatedly fetching data and maximizes data access locality whenever possible.
%RPCs are also used in RDMA-enabled graph databases because of their better data locality compared to repeated RDMA fetches ~\cite{A1}.

\subsection{Our Approach}
\label{sec:approach}
\spara{Latency-bound replication problem}
This paper formalizes the latency-bound replication problem and proposes algorithms to solve it.
The problem takes a workload model, a system model, and a latency  bound as input and produces a \emph{replication scheme}, indicating which objects need to be replicated on which server to respect the latency bound with minimal cost.
We show that the latency-bound replication problem is NP-hard.

Optimizing query latency is one of the traditional goals of distributed query optimization (see e.g.~\cite{kossmann2000state}).
However, the two problems are fundamentally different.
Distributed query optimization aims at finding a physical query execution plan that minimizes query latency when the data placement scheme, which describes how data is partitioned and potentially replicated, is given.
A query optimizer does not change the data placement so it cannot, in general, ensure arbitrary upper bounds on the latency of arbitrary queries.

This work considers a different problem: how to change the data placement to ensure an upper bound on latency given a physical query execution plan.
First, we consider the plan in a configuration when there is \emph{no replication} and model the causality and locality of data accesses.
This model allows the replication algorithm to reason about the latency of the query.
Our replication algorithm then evaluates the effect on latency of replicating data in the original plan by replacing distributed traversals with local accesses.

% The system can store replica objects in the data store together with the master objects.
% %Replica objects create potential implications for subquery routing.
% %The subquery routing component must now locate the right destination to respect the query's latency constraint.
% In principle, the subquery routing component could now keep per-query mappings of which remote copy of an object (master or replica) each query must access to respect its latency bound.
% This would result in large routing tables and complex routing logic.
% In this paper, we forgo this complexity and adopt the simplest possible routing logic, which we call \emph{thrifty routing}: when a query accesses an object, it accesses a \emph{local} master or replica copy of that object if available, otherwise we use the unmodified sharding-based subquery routing component to access the master copy of that object, as done in the original system without replication (see Figure~\ref{fig:overview-diagram}).
% This approach is thrifty because it does not require any additional routing information and it is easier to adopt in existing systems.
% We show that, even with this simple thrifty routing logic, the latency-bound replication problem is NP-hard.

\spara{Greedy replication algorithm}
We propose a family of greedy algorithms that incrementally add replica objects after analyzing one path of one causal access path at a time.
Showing that these incremental additions must respect the required latency bounds is surprisingly non-trivial: adding replicas to optimize one path might actually violate the latency constraint for some path that was optimized previously.
We introduce the notion of \emph{latency-robustness}  and show that if a replication scheme is latency-robust for a path then any arbitrary modification of it will still preserve the latency bound for that path.
We finally propose a greedy algorithm that meets these constraints and minimizes the replication cost.

\spara{Handling reconfigurations and failures}
Query execution systems may need to change the set of servers they run on or handle server failures.
Preserving the latency constraints after these events may require updating the replication scheme.
We propose extensions of our algorithm that incrementally update the replication scheme without analyzing the entire workload from scratch.

\section{Latency-Bound Replication Problem}
\label{sec:problem-def}

In this section, we formalize the latency-bound replication problem and show that it is NP-hard.

\spara{Workload and system model}
We start by formalizing the the workload and system models described in Section~\ref{sec:overview}.

We model a \emph{dataset} $D$ as a set of abstract \emph{objects}.
The storage cost of an object $v$ is defined by the storage function $f(v)$.
We model a \emph{workload} $W$ as a set of queries.
Executing a query involves accessing multiple objects in the dataset.
If the access to an object $v$ \emph{causally precedes} the access to another object $u$ \emph{in the same query}, we say that $v$ happens before $u$, or $hb(v\rightarrow u)$.
The happen before property is transitive, that is, $hb(v \rightarrow u)$ and $hb(u \rightarrow z)$ implies $hb(v \rightarrow z)$.
To model causally-related data accesses of a query, we introduce the notion of \emph{causal access path}, which we define as follows, and a query $Q$ as the set of all its causal access paths.
\begin{definition}[Causal access path]
    A causal access path $p$ over a dataset $D$ is a path where each node is mapped to an object in $D$.
    A node $v_p$ is the parent of another node $v_c$ if and only if $hb(v_p \rightarrow v_c)$ and there is no object $u \in D$ such that $hb(v_p \rightarrow u)$ and $hb(u \rightarrow v_c)$.
\end{definition}

The workload is executed by a distributed system consisting of a \emph{set of servers} $S$.
Each server stores a partition of the dataset $D$ determined by a \emph{sharding function} $d$, which maps each object of $D$ to one server in $S$.
Each server has a maximum storage capacity $M_s$

\begin{table}[t]
\begin{footnotesize}
\begin{tabular}{|l|l|}
\hline 
\multicolumn{2}{|c|}{\textbf{Inputs: Workload model}} \\ 
\hline 
$D$ & Dataset: set of objects constituting the dataset. \\ 
\hline 
$f(v)$ & Storage cost of object $v$.\\
\hline
$W$ & Workload: set of queries.\\
\hline
$Q$ & Query: set of causal access paths.\\
\hline 
$t_Q$ & Latency constraint of a query $Q$.\\
\hline
\multicolumn{2}{|c|}{\textbf{Inputs: System model}} \\ 
\hline 
$S$ & Set of servers in the system. \\ 
\hline 
$d(v)$ & Sharding function: server with original copy of object $v$.\\
\hline 
$\rho$ & Access function.\\
\hline
$M_s$ & Maximum storage capacity of server $s$.\\
\hline
\multicolumn{2}{|c|}{\textbf{Output}} \\ 
\hline 
$r(v)$ & Replication scheme: location of all copies of object $v$.\\
\hline 
\multicolumn{2}{|c|}{\textbf{Auxiliary notation}} \\ 
\hline 
$h(p,r,\rho)$ & Latency of a path $p$ under $r$ and $\rho$.\\
\hline 
$l_Q$ & Latency of a query $Q$.\\
\hline 
$f_r(s)$ & Storage cost of a server $s$ under the replication scheme $r$.\\
\hline 
\end{tabular} 

\end{footnotesize}
\caption{Notation}
\end{table}

A \emph{replication scheme} $r$ maps each object $v$ to a set of servers, one of which is the original location of the object according to the sharding function.
Initially, before we run our algorithms, there are only the original copies of objects in the system and $r(v)=d(v)$ for each object $v$.
As we add replicas for an object $v$, we have that $d(v) \in r(v)$.
%
%\begin{definition}[Replication scheme]
%    A replication scheme $r$ extending a sharding function $d$ is a function $r: D \rightarrow 2^S$ such that $\forall v \in D, d(v) \in r(v)$.
%\end{definition}

When a query accesses a data object $v$ in the original system configuration with no replication, it does so at the server that holds $v$ according to the sharding function.
With replication, the query can avoid distributed traversals if a local replica is available.
Formally, let $p$ be a causal access path and $r$ a replication scheme.
The access function $\rho$ associates each non-root node $v$ in $p$ with a server $\rho(r,v) \in r(v)$. 
\begin{equation}
    \label{eqn:constrained}
\rho(r,v) = \left\{
\begin{array}{rl}
    \rho(r,u) & \text{if } \exists u \text{ parent of } v \text{ in } p \text{ s.t. } \rho(r,u) \in r(v),\\
    d(v) & \text{otherwise}.
\end{array} \right.       
\end{equation}    
%% COMMENT === v must be non-root because in the NP-hardness proof we need to route the root node of a causal access tree to a replica. This is also reflected in the definition of upward replication and the proof of optimality. We don't want to constraint the initial routing of a query.
%With thrifty routing, queries access \emph{local} master or replica objects if available. 
%If not, the query executor issues a subquery RPC using sharding-based routing.
%We will show that even with this simple routing function the problem is NP-hard and proving sufficient correctness conditions for greedy algorithms is highly non-trivial.

\spara{Problem definition}
We use a \emph{latency function} $l$ to count the number of distributed traversals in a path.
Given two servers $s_1,s_2$ where two consecutive accesses in the path occur, we have that $l(s_1, s_2)$ is equal to $1$ if $s_1 \neq s_2$, or $0$ otherwise. 
The servers where vertices are accessed is determined by the access function. 

\begin{definition}[Latency cost of a path]
    Let $D$ be a dataset, $p = \langle v_{p_1}, \ldots,  v_{p_n}\rangle$ a path over objects in $D$, and $r$ a replication scheme over a set of servers $S$.
    The latency cost $h(p)$ of $p$ is defined as:
    \begin{equation}
        h(p,r,\rho) = \sum_{(v_{p_i},v_{p_{i+1}}) \in p} l(\rho(r,v_{p_i}),\rho(r,v_{p_{i+1}})) 
        \label{eqn:latency-subpath-repl}
    \end{equation}
\end{definition}

\begin{definition}[Latency cost of a query]
    Let $Q = \{p_1, \ldots, p_m \}$ be a query over a dataset $D$ and $r$ a replication scheme.
    The latency cost $l_Q$ of $Q$ under $r$ is defined as:

\begin{equation}
    l_Q = \max_{C_i\in Q} \Big(\max_{p \in Q} h(p,r,\rho) \Big)
    \label{eqn:latency-cost-repl}
\end{equation} 
\end{definition}

We now formally define our problem in Definition~\ref{def:prob-statement}.
Our goal is to find a replication scheme $r$ that guarantees all queries complete within their latency constraints (first constraint) and the storage costs among all servers are balanced (second constraint) while the total replication cost is minimized (minimization problem). 
We use $ f_r(s) = \sum_{v \in D:\: s \in r(v)} f(v)$ to denote the storage cost of a server $s$ under a replication scheme $r$.

\begin{definition}[Latency-bound replication problem]\label{def:prob-statement}
    Let $D$ be a dataset, $W = \{Q_1, \ldots, Q_m\}$ a workload consisting of a set of queries, each associated with a latency constraint $t_{Q_i}$, $S$ a set of servers, $d$ a sharding function, $l$ a latency function, $f$ a storage cost function, $M_s$ the storage capacity of a server $s$, and $\epsilon$ a load imbalance constraint.
    The latency-bound replication problem asks for a replication scheme $r$ extending $d$ such that:
    \begin{equation}
        \begin{aligned}
            \min_r \quad & \sum_{s \in S} f_r(s) \\
            \textrm{s.t.} \quad & \forall Q_i \in W: \: l_{Q_i} \leq t_{Q_i} \quad \wedge\\
            & \forall s, s' \in S:\: |f_r(s) - f_r(s')| \leq \epsilon   \\
            & \forall s \in S, f_r(s) \leq M_s\\
        \end{aligned}
        \label{eqn:problem}
    \end{equation}
\end{definition}

We now show that this problem is NP-hard. It is NP-hard even in the special case where we disregard the load imbalance constraint, and only need to check whether there is a solution satisfying the latency constraint and the storage capacity constraint. %  in the special case we consider in this paper, where the routing function is constrained to be a thrifty routing function.

\begin{theorem}\label{thm:NP-hardness}
    The latency-bound replication problem is NP-hard. Moreover, it is NP-hard to check whether there exists a replication scheme satisfying a given latency bound and storage capacity constraint for each server.
\end{theorem}
\begin{proof}
	(Sketch) We show that checking that there exists a replication scheme satisfying a given latency bound and storage capacity constraint for each server, called \emph{the latency-storage feasible problem}, is NP-hard. This implies NP-hardness of the latency-bound replication problem.   The proof has two steps. Our key insight is in the first step, where we introduce the \emph{min-bridge bisection problem}: given a graph, the task is to partition the graph into two subgraphs with an equal number of vertices such that the number of vertices with neighbors in the other partition is bounded. We then reduce the latency-storage feasible problem to the min-bridge bisection problem. 
	%In the first step, we reduce the latency-storage feasible problem to a problem we call \emph{min-bridge bisection problem}, where we want to partition the graph into two subgraphs with an equal number of vertices such that the number of vertices with neighbors in the other partition is bounded.
	In the second step, we reduce that problem to the min-bisection problem for 3-regular graphs, which is NP-hard~\cite{bui1987graph}. The two reductions imply that the latency-storage feasible problem is NP-hard.
The full proof is in the Appendix.
\end{proof}

\section{Greedy Algorithms}
\label{sec:greedy}
Since finding an optimal solution the latency-bound replication problem, and even checking that there exists a feasible solution, are NP-hard by Theorem~\ref{thm:NP-hardness}, we devise a family of greedy algorithms that always guarantee the latency constraint while optimizing the replication cost as much as possible and checking if the other constraints are preserved.
Our experiments showed that the replication schemes found by the greedy algorithm have small storage cost and small imbalance parameter ($\epsilon = 0.02$ in our experiments). 
Herein, we refer to a solution that satisfies the latency constraint as a \emph{latency-feasible solution}.
We also introduce a sufficient condition to find latency-feasible solutions called \emph{latency-robustness}.

\begin{algorithm}[t]
    \caption{Greedy Latency-Bound Replication Algorithm}
\label{algo:generic}
\centering
\begin{small}
\begin{algorithmic}[1]
\State $r_0 \gets d$
\State $i \gets 0$
\ForAll{$Q \in W$}
    \ForAll{path $p \in Q$}
        \State $i++$
        \State $r_i \gets$ \textsc{update}$(r_{i-1}, p)$ %\Comment{$\forall v \in V, r_{i-1}(v)\subseteq r_i(v)$}
    \EndFor
\EndFor
\State \Return $r_{i}$
\end{algorithmic}
\end{small}
\end{algorithm}

\subsection{A Family of Greedy Algorithms}
A generic greedy latency-bound replication algorithm is shown in Algorithm~\ref{algo:generic}.
It starts from a replication schemes $r_0$ with no replicas and only original objects, placed according to the sharding function provided as input. Recall that the latency of a query is defined as the latency along its slowest root-to-leaf path (Eqn.~\ref{eqn:latency-cost-repl}).
To bound the latency of a query, we can thus iterate across all paths and bound the latency of each of them.
Given a path $p$, the \textsc{update} function inputs the current replication scheme $r_{i-1}$ and the path $p$.
It outputs a new replication scheme $r_i$ that adds replicas to $r_{i-1}$ to satisfy the user-defined latency constraint over the path $p$.
Each path is visited only once, so the algorithm can materialize only one path at a time using the workload analyzer.

There can be many possible implementations of the \textsc{update} function.
In the following, we discuss conditions on the output of the \textsc{update} function that ensure the correctness of Algorithm~\ref{algo:generic}.
We then propose a specific implementation of the \textsc{update} function that meets these correctness conditions.

\begin{figure*}
\includegraphics[width=\textwidth]{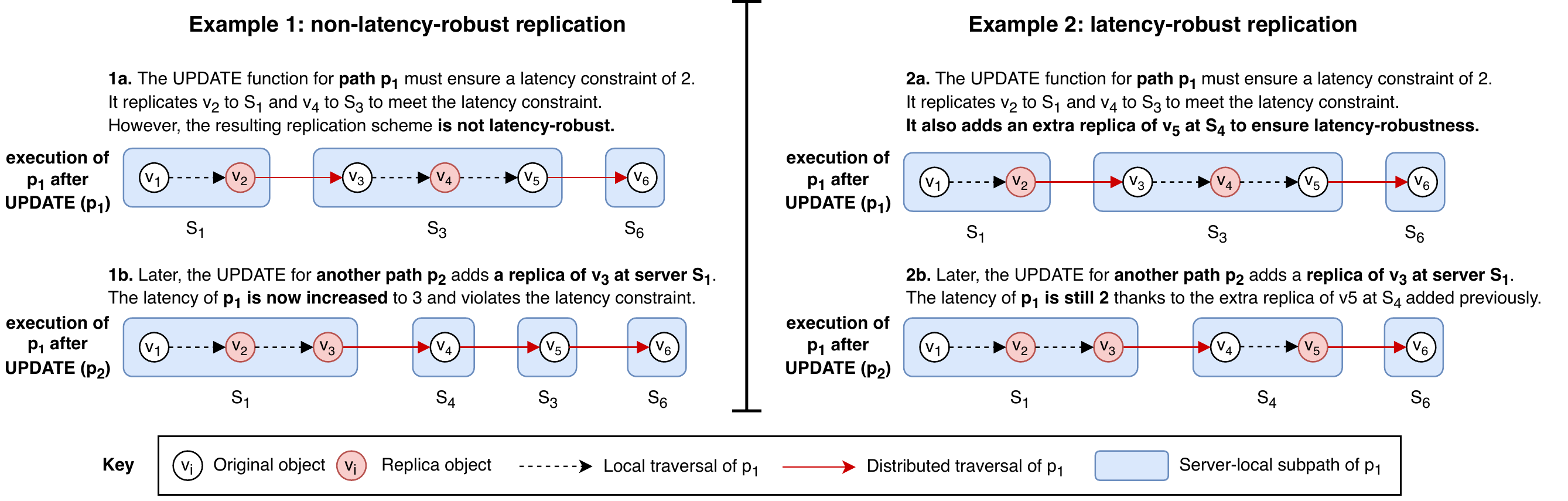} 
\caption{An example motivating latency-robustness.}
\label{fig:robust}
\end{figure*}

\subsection{Correctness Condition}
%To generate feasible solutions to the latency-bound replication problem, Algorithm~\ref{algo:generic} must output replication schemes that respect the constraints defined in Equation~\ref{eqn:problem}.
The \textsc{update} function can only see one path of the workload at a time.
An implementation of that function can easily ensure that the latency constraint is respected on the path under exam.
We argue, however, that respecting the latency constraints on one path at a time \emph{is not sufficient} to ensure that Algorithm~\ref{algo:generic} produces latency-feasible solutions.
We then propose an additional condition called \emph{latency-robustness} and show that it is sufficient.

\spara{Motivation for latency-robustness}
We now show a run of the greedy algorithm where the \textsc{update} function produces a replication scheme that respects the latency constraint on a path $p_1$.
Later, it adds new replicas to reduce the latency of another path $p_2$ and these new replicas increase the latency of $p_1$.

Figure~\ref{fig:robust}(1a) shows the replication scheme after the first path $p_1$ is processed by the \textsc{update} function.
The path accesses six objects, from $v_1$ to $v_6$, in a sequence.
%The sharding function provided as input assigns the master copy of $v_3$ and $v_5$ to server $s_3$, while the master copy of the other objects are all assigned to different servers.
Suppose that the latency bound specified by the problem definition is $2$ distributed traversals. 
%Without replication, the latency of the path is $5$ distributed traversals.
%The \textsc{update} function produces the replication scheme depicted as Example 1a, where replicas of $v_2$ and $v_4$ are added to $s_1$ and $s_3$ respectively. 
The replication scheme shown in Figure~\ref{fig:robust}(1a) ensures that path $p_1$ respects the latency constraint of the problem.
Nonetheless,  later additions of replica objects can violate the latency constraint on path $p_1$.

Suppose that another path $p_2$ (not depicted in Figure~\ref{fig:robust}) is later processed by the \textsc{update} function and assume that ensuring the latency constraint for that path requires adding a replica of $v_3$ to server $s_1$.
This changes the execution of path $p_1$ as shown in Figure~\ref{fig:robust}(1b).
The query executor at $s_1$ now accesses a local replica copy of $v_3$ rather than executing a distributed traversal to access the original copy of $v_3$ at $s_3$, as it did with the previous replication scheme.
Path $p_1$ now misses the opportunity of accessing $v_3$, $v_4$, and $v_5$ locally at $s_3$.
Instead, after accessing the local replica of $v_3$ at $s_1$, the path accesses the original copy of $v_4$, which we assume is located at $s_4$, because a replica of $v_4$ is not available at $s_1$.
The following traversal is also distributed because $s_4$ has no local replica of $v_5$, so we need again to access the original copy.
The latency of $p_1$ now increases to $3$ and violates the latency constraint.

\spara{Latency robustness}
We introduce a property of replication schemes called \emph{latency robustness} and show that, if enforced by the \textsc{update} function, it avoids the problem we have discussed.
Informally, a replication scheme is latency robust for a path $p$ if each object in a server-local subpath of $p$ is replicated to all servers storing the original copies of all its predecessors in the subpath.
A server-local subpath is a sequence of consecutive accesses in a path that is local to some server; see Figure~\ref{fig:robust} and Definition~\ref{def:local-subpath} for a formal definition.
The first access in the subpath is a distributed traversal and is directed to the server storing the original copy of the object being accessed.
The subsequent accesses in the subpath are all local to the same server.

Consider again the example of Figure~\ref{fig:robust}(1a) and in particular the server-local subpath at server $s_3$.
Latency-robustness is violated for object $v_5$, which has two predecessors, $v_3$ and $v_4$.
Latency-robustness requires that a replica copy of $v_5$ be kept by the server holding the original copy of $v_4$, which is $s_4$ (see Figure~\ref{fig:robust}(1b)).
However, the \textsc{update} function violates this requirement because it does not add that copy, and this leads to a latency violation, as we discussed previously. 

Latency robustness avoids this problem.
Consider now the example of Figure~\ref{fig:robust}(2a).
To ensure latency robustness, the \textsc{update} function now replicates $v_5$ to server $s_4$.
Now, even if we add a replica of $v_3$ to $s_1$, the path still only requires $2$ distributed traversals (see Figure ~\ref{fig:robust}(2b)).
This is because when the \textsc{update} function processes $p_1$ it has added enough replicas to deal with any possible later updates to the execution of $p_1$.
Even if some accesses that are currently in a server-local subpath are moved to another server, we show that the number of distributed traversals required to perform the accesses in the subpath will not increase.

\spara{Correctness proof}
We now show that if an implementation of the \textsc{update} function produces latency-feasible replication schemes on its input path $p$ \emph{and is latency-robust for $p$} then any arbitrary addition of replica objects to $r$ will be latency-feasible for $p$.
We start by formalizing the notion of server-local subpath.

\begin{definition}[Server-local subpath]\label{def:local-subpath}
    Let $p:\langle v_{p_1}, v_{p_2}, ..., v_{p_n}\rangle$ be a root-to-leaf causal access path of a query and $r$ a replication scheme.
    A server-local subpath $g_{p,r}^i$ of $p$ under $r$ for $i$ is a subpath $g_{p,r}^i = \langle v_{p_j}, ..., v_{p_k}\rangle$  of $p$ such that $h(g_{p,r}^i, r, \rho) = 0$ and:
    \begin{itemize}
        \item if $j > 1$, then it holds that $h(\langle v_{p_1}, \ldots, v_{p_{j-1}} \rangle, r, \rho) = i$ and $\rho(r,v_{p_{j-1}}) \neq \rho(r,v_{p_j})$, and
        \item if $k < n$, then it holds that $\rho(r,v_{p_k}) \neq \rho(r,v_{p_{k+1}})$.  
    \end{itemize}
\end{definition}
%All accesses in the server-local subpath $g_{p,r}^i$ are local to the server $d(g_{p,r}^i[0])$, which stores the master copy of the first object accessed in the subpath. 

We can now define latency-robustness as follows.

\begin{definition}[Latency-robustness]
    Let $p = \langle v_{p_1}, v_{p_2}, ..., v_{p_n}\rangle$ be a root-to-leaf path of a query and $r$ a replication scheme.
    The replication scheme $r$ is robust for $p$ if and only if for each server-local subpath $g_{p,r}^i = \langle v_{p_j}, ..., v_{p_k}\rangle$ of $p$, it holds that:
	\begin{equation}
    	 \forall x,y \in[j,k], x < y : d(v_{p_x}) \in r(v_{p_y}).
	\end{equation} 
\end{definition}

We can now prove that the greedy latency-bound replication algorithm (Algorithm~\ref{algo:generic}) outputs a latency-feasible solution if the \textsc{update} function is latency-feasible and latency-robust for path $p$.
The proof is in the Appendix.
%We finally prove that a correctness condition for the greedy latency-bound replication algorithm (Algorithm~\ref{algo:generic}), which depends on how the \textsc{update} function is implemented.

\begin{theorem}
\label{thm:correct-update}
The greedy latency-bound replication algorithm (Algorithm~\ref{algo:generic}) returns a latency-feasible solution for the latency-bound replication problem if any invocation of \textsc{update($r_{i-1}$, $p$)} returns a replication scheme $r_i$ that is (i) latency-robust for $p$, and (ii) it is latency-feasible for workload consisting only of $p$.
\end{theorem}

%\hl{Feasible in the above theorem just mean the latency-bound is satisfied, not the balanced constraint or maximum storage constraint. }

\vspace{-0.5em}
\subsection{The Full Replication Algorithm}
\label{sec:rep-algo}
We now present an algorithm that instantiates the \textsc{update} function and respects the correctness conditions of Theorem~\ref{thm:correct-update}.

\begin{comment}
\begin{figure}[t!]
    \centering
    \begin{subfigure}[b]{0.3\linewidth}
        \centering
        \includegraphics[width=\linewidth]{figures/upward_downward_G.pdf}
        \caption{Dataset graph.}
        \label{fig:up_down_G}
    \end{subfigure}
    \hfill
    \begin{subfigure}[b]{0.3\linewidth}
        \centering
        \includegraphics[width=\linewidth]{figures/upward_replication.pdf}
        \caption{Thrifty routing + upward replication.}
        \label{fig:upward_replication}
    \end{subfigure}
    \hfill
    \begin{subfigure}[b]{0.31\linewidth}
        \centering
        \includegraphics[width=\linewidth]{figures/downward_replication.pdf}
        \caption{Routing with arbitrary replication.}
        \label{fig:downward_replication}
    \end{subfigure}
       \caption{Examples of possible routing paths (dashed) under different replication and routing schemes. Replicas are represented in solid colors. }
       \label{fig:upward_downward_replication}
\end{figure}
\end{comment}

\spara{Upward replication}
Our algorithm only considers a class of replication schemes that only replicate an object at the server where its predecessor object in a path is accessed.
This is because a query accessing data at a server will either access a local data replica of the object or the original copy.
Therefore, replica objects that do not follow upward replication are never accessed because they are not colocated with any parent of any path.

\begin{definition}[Upward replication scheme]
    Let $D$ be a dataset and $W = \{ Q_1, \ldots, Q_m\}$ be a workload.
    A replication scheme $r$ is an upward replication scheme if the following holds: if $\rho(r,v) \neq d(v)$ for some causal access path $p$ of a query $Q_i\in W$ and non-root node $v$ of $p$ then $\rho(r,u) = \rho(r,v)$, where $u$ is the parent of $v$ in $p$.
    \label{def:upward_replication}
\end{definition}

\begin{comment}
An example of an upward replication function is shown in Figure~\ref{fig:upward_downward_replication}.
Consider accessing the objects $\langle v_2, v_4, v_6\rangle$ in sequence and suppose that we can only perform one distributed traversal.
With upward replication, we can replicate $v_6$ to the server where its predecessor $v_4$ is accessed in path, which is $s_2$ (see Figure~\ref{fig:upward_replication}).
This ensures that the local replica of $v_6$ can be accessed using thrifty routing.
Doing the opposite, which is replicating $v_4$ to be co-located with $v_6$ would not help with thrifty routing (see Figure~\ref{fig:downward_replication}).
After accessing $v_2$ at server $s_1$, a server would need to locate the replica of $v_4$ at server $s_3$. 
This would require extra routing information, since the default sharding function only indicates the location of the original copies of objects.
Storing and querying this extra routing data can become expensive, since different paths might require accessing different replicas of the same object.
These problems are avoided by the use of a thrifty routing scheme.
\ms{This could go}
\end{comment}

We prove that any optimal solution to the latency-bound replication problem must be an upward replication function. 
The proof of the theorem is in the Appendix.

\begin{theorem}
    \label{thm:upward}
    Any optimal replication scheme solving the latency-bound replication problem is an upward replication scheme. 
\end{theorem}

\spara{Replication algorithm}
We now propose an algorithm implementing the \textsc{update} function that searches for upward replication schemes that ensure the latency bound on $p$ with minimal replication cost.
%It is based on the observation that the number of distributed traversals of a path $p$ is equal to the number of its server-local subpaths minus one. 
%If this exceeds the latency bound,  the algorithm \emph{selects} a subset of these subpaths and \emph{merges} the non-selected subpaths to the selected ones.
%Merging is done by replicating the objects of each non-selected subpath to the server of the selected subpath preceding it.
%
The pseudocode of the replication algorithm is in Algorithm~\ref{algo:proposed}. 
For simplicity, we consider paths where there is only one barrier to collect the results.
The algorithm starts by enumerating the set of server-local subpaths of the path under the original sharding scheme $d$, without considering replicas (Line \ref{algo3:enum_begin}-\ref{algo3:enum_end}).
\setlength{\textfloatsep}{0.5em}
\begin{algorithm}[t]
    \caption{Replication Algorithm}\label{algo:proposed}
\begin{small}
\begin{algorithmic}[1]
    \Procedure{update}{$r, p$}
%        \State $n \gets$ length($p$)  \Comment{$p= \langle v_{p_1}, \ldots, v_{p_n}\rangle$}
        %\State $g_p^0 \gets \langle v_{p_1} \rangle$
        %\For{$h \gets 1$ to $n$}
        %    $g_p^h \gets \langle \rangle$
        %\EndFor
        \State $G_{p,d} \gets$ set of server-local traversal subpaths of $p$ under $d$ \label{algo3:enum_begin}
        \State $h \gets |G_{p,d}| -1 $ \label{algo3:enum_end}
        %\For{$i \gets 1$ to $n-1$}
        %    \If{$d(v_{p_i}) \neq d(v_{p_{i+1}})$}
        %        $h = h + 1$
        %    \EndIf
        %    \State append$(g_{p,r}^h, v_{p_{i+1}})$ 
        %\EndFor

        \If{$h > t_{Q_i}$}  \label{algo3:if_violates}
            \State $\emph{Candidates} \gets$ set of subsets of $ \{1,...,h\}$ of size $t_{Q_i}$ \label{algo3:enumerate}
            \For {$\Delta \in \emph{Candidates}}$ 
 \label{algo3:for_each_plan}
                \State $\Delta \gets \Delta \cup \{0\}$ \label{algo3:retain_0}
                \State $\Delta.cost \gets 0$ 
                \State $\Delta.r \gets r$
                \State{\emph{// Non-selected subpaths: replicate their objects}}
                \For{$i \gets [1,h]$ s.t. $i \not\in \Delta$ } \label{algo3:each_subpath}
		                \State{\emph{// Preceding selected subpath}}
                        \State $j \gets$ largest value in $\Delta$ smaller than $i$ \label{algo3:closest_parent}
                        \For{$v \in g_{p,d}^i$} \label{algo3:for_each_v}
                            \For{$k \gets j$ to $i - 1$} \label{algo3:robust} \Comment{\emph{Latency-robustness}}
                            	\For{$u \in g_{p,d}^k $}
	                                \If{$d(u) \not\in \Delta.r(v)$}
    	                                \State $\Delta.r(v) \gets \Delta.r(v) \cup \{d(u)\}$ \label{algo3:add_replica}
        	                            \State $\Delta.cost \gets \Delta.Cost + f(v)$ \label{algo3:add_cost}
            	                    \EndIf
                            	\EndFor
                            \EndFor
                        \EndFor
                \EndFor
	            \If{$\Delta.r$ violates storage capacity or load balance}
    	        	\State remove $\Delta$ from \emph{Candidates} \label{algo3:feasible}
        	    \EndIf
            \EndFor
        \EndIf
        \If{\emph{Candidates} $\neq \emptyset$} \label{algo3:pick_solution}
        	\State \Return $\Delta.r \in $ \emph{Candidates} with minimal $\Delta.cost$
        \Else
	        \State \Return \emph{no-solution-found} \label{algo3:no_solution}
        \EndIf
    \EndProcedure
\end{algorithmic}
\end{small}

\end{algorithm}
%The algorithm first constructs the traversal subpaths of $p$ (Line \ref{algo3:make_subpaths}).
A path $p$ violates its latency constraint if it has more than $t_{Q_i} + 1$ server-local subpaths.
If that is the case (Line \ref{algo3:if_violates}), the algorithm \emph{selects} a set of $t_{Q_i} + 1$ subpaths to retain and \emph{merges} each non-selected subpaths to the selected subpath preceding it.
Merging uses upward replication and incurs a replication cost.
To minimize this cost, the algorithm enumerates all candidate sets of subpaths to select and picks the candidate set with the lowest merging cost, considering that some object may have already been replicated in the current replication scheme $r$.

More specifically, the algorithm exhaustively enumerates the indices of all candidate sets of $t_{Q_i}$ subpaths (Line \ref{algo3:enumerate}) and adds to each candidate set the first subpath (Line \ref{algo3:retain_0}), since the algorithm assumes that the first data access of a query is routed according to the original sharding scheme $d$.
For each candidate $\Delta$, the algorithm updates the replication scheme $r$ to merge the non-selected subpaths and finds the associated replication cost (Line \ref{algo3:for_each_plan}-\ref{algo3:feasible}). 

Given a candidate $\Delta$ and the index $i$ of a subpath  that is not selected in $\Delta$ (Line \ref{algo3:each_subpath}), the algorithm merges the subpath to the preceding server-local subpath, which has index $j$.
To do so, it replicates each object of the non-selected subpath $g^i_{p,d}$ to the server where the accesses of the selected subpath occur in the candidate replication scheme.
To ensure latency robustness, it is also necessary to replicate the object to the servers storing the the original copies of the other preceding objects $u$ in the subpath.
That is why the algorithm iterates over all such subpaths, whith indices $k \in [j,i-1]$, and replicates $v$ to $d(u)$ for each $u \in g^k_{p,d}$.
Note that the resulting replication scheme is an upward replication scheme, because each replicated object is co-located with its predecessor in the path.

The algorithm then removes candidates that do not respect the storage capacity and load balancing constraints of the latency-bound replication problem.
Note that even checking that there exists a feasible solution satisfying the latency and storage capacity constraints is NP-hard as per Theorem~\ref{thm:NP-hardness}.
Finally, the algorithm returns the replication scheme with the lowest cost (Line \ref{algo3:pick_solution}-\ref{algo3:no_solution}).
If no candidate satisfies the constraints, the algorithm returns a flag indicating the algorithm cannot find a feasible solution.

It is easy to see that the algorithm respects the correctness conditions of Theorem~\ref{thm:correct-update}.
Latency-robustness is ensured by construction.
The feasibility of the solution on $p$ is checked explicitly on each replication scheme.
Note that if any of the candidates produces a feasible replication scheme for $p$, the algorithm will find it, since it exhaustively explores all candidates.

%Note that if $p$ satisfies the latency constraint under $r$ using the thrifty routing function, but not under $d$, and $r$ is robust for $p$, it means that one of the possible replication plans will have a replication cost equal 0, and will be chosen by the algorithm as it has the minimum cost. 
%No replication is needed for $p$ in this case. 

\spara{Performance optimizations}
The number of candidates equals ${n-1 \choose  t_{Q_i} - 1}$ for a path with $n$ accesses, which is relatively small for low-latency queries. %\hl{can we use dynamic programming?}\ms{We discussed it with Nathan and he concluded that it was not a good idea, but I don't remember why.}
To improve performance, our implementation of the algorithm iterates over the set candidates twice. 
In the first pass, the algorithm only computes the cost of the replication scheme associated to each candidate.
In the second pass, it iterates over the candidates again in ascending cost order, computes the replication scheme of each candidate, and checks it if it is feasible.

%\begin{figure}
%	\includegraphics[scale=.8]{figures/local_min_example.pdf} 
%	\caption{Pruning redundant paths}
%	\label{fig:short_q_7}
%\end{figure}

Our algorithm can also prune some paths.
If two paths have roots accesses occurring at the same server and are identical except from their root, then any replication scheme that is feasible for one path is feasible also for the other so we can process only one path.
%Consider the two causal access trees $C$ and $C'$ of a query in Figure \ref{fig:short_q_7}. 
%Since the two trees share a common subpath $\langle v_4,v_5 \rangle$ and the master of their root object is stored at the same server $s_2$, we can process only the path in one of the trees. 
With this pruning technique, the number of root-to-leaf paths set $P$ of a query can be reduced by at most a factor of $|S|$. 
%In our implementation, we keep track of each pair of the root server and checked $2^{nd}$ hop vertices for each query, and prune if the $\langle root\ server,2^{nd}\ hop\ vertex\rangle$ pair has been optimized previously. 
%We present our proposed replication algorithm in Algo \ref{algo:proposed}. 
%In Algo \ref{algo:proposed}, we maintain a set of already checked servers and subpath pairs for a given match (Line 3). 
%For each root-to-leaf path $p$ (Line 5), we first check whether the path can be pruned (Line 6).
%If not, the algorithm constructs $m_p$ (Line 7-13), enumerates all possible incremental-robust plans (Line 14), compute the replication cost of each plan using the proposed cost model (Line 16-27), and replicates according to the replication plan with the minimum cost (Line 27-33).
%Finally, the algorithm adds the newly checked server and subpath pair to the set $CS$ (Line 34), and loop until all matches are optimized.  

% Note that this pruning technique only works for queries where the root node and the first hop have a many-to-one or many-to-many relationship. 
% This is because this technique prunes the redundant computation. 
% However, for queries where the first two hops have a one-to-one or one-to-many relationship, each first-hop vertex of the query will only be included in one root-to-leaf path; hence no redundant computation can be pruned. 

\spara{Workload analysis}
%The algorithm we described takes as input a workload model expressed in terms of a set of causal access paths.
We implemented workload analyzers that take a dataset and a set of query types as input and enumerate all the paths in the workload.
Its output can be an \emph{overapproximation}: it only has to  \emph{include} all the paths that actually occur in the workload.
The greedy algorithm materializes only the paths currently processed  by the \textsc{update} function.

\subsection{Updating the Replication Scheme}
\label{sec:rep-algo-dyna}
% We now discuss scenarios where it is helpful to incrementally update the replication scheme without rerunning the replication algorithm on the entire workload.

To change the set of servers in the system, handle server faults, or change the sharding function, the query execution system might \emph{reshard} the dataset, i.e., relocate the original copies of some objects.
We show that we can extend our replication algorithm to update the replication scheme incrementally when reshards occur.
Replication for fault tolerance is an orthogonal issue handled by the query execution system. 

First, we extend Algorithm~\ref{algo:proposed} such that together with a replication scheme, the \textsc{update} function also incrementally adds to a data structure called the \emph{resharding map} $RM$.
Every time we add a replica of an object $v$ to the server storing the original copy of an object $u$ (Line~\ref{algo3:add_replica} of Algorithm~\ref{algo:proposed}), the algorithm adds a mapping $\langle u, v \rangle$ to $RM$.
The final colocation map is returned as an output of the algorithm and can be looked up by the query execution system.
The algorithm also outputs an \emph{reference count} $RC(v,s)$ for each replica object $v$ and server $s$, which counts how many distinct original objects sharded to $s$ the replica object $v$ is associated to.

Next, we run an incremental algorithm when the query execution system starts resharding.
The system indicates which original objects must be transferred to which server to our algorithm, which  queries the resharding map and transfers the associated replica objects.
For each object $u$ that is resharded from a server $s$ to a different server $s'$, the new replication scheme must transfer to $s'$ a copy of all objects $v$ such that $\langle u, v \rangle \in RM$, unless a original or replica copy of $v$ is already present at $s'$.
Our algorithm then increases $RC(v,s')$, decreases $RC(v,s)$, and deletes the replica object $v$ from $s$ if the count is smaller than one.
The resulting replication scheme still respects the constraints of the latency-bound replication problem and is latency-robust.
This is because Algorithm~\ref{algo:proposed} operates by co-locating object replicas with the original copy of other objects, regardless of how the sharding scheme places those original copies.

\section{Evaluation}
\label{sec:eval}
\begin{figure*}
  \begin{subfigure}[b]{.156\linewidth}
    \centering
    \includegraphics[width=\textwidth]{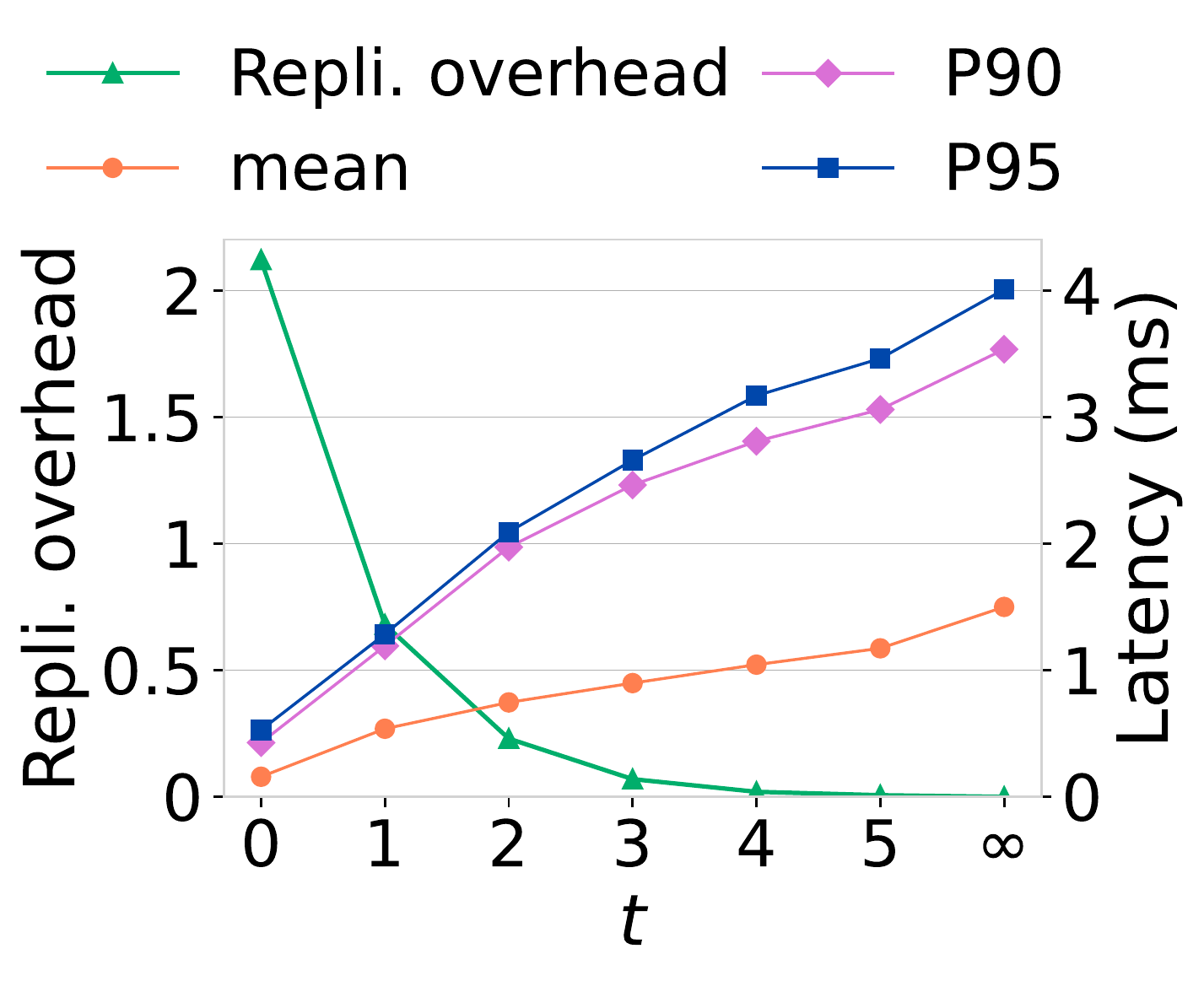}
    \caption{SNB: latency\\ vs. replication}
    \label{fig:short_q_latency_not_normalized}
  \end{subfigure}\hspace{0.1mm}
  \begin{subfigure}[b]{.161\linewidth}
    \centering
    \includegraphics[width=\textwidth]{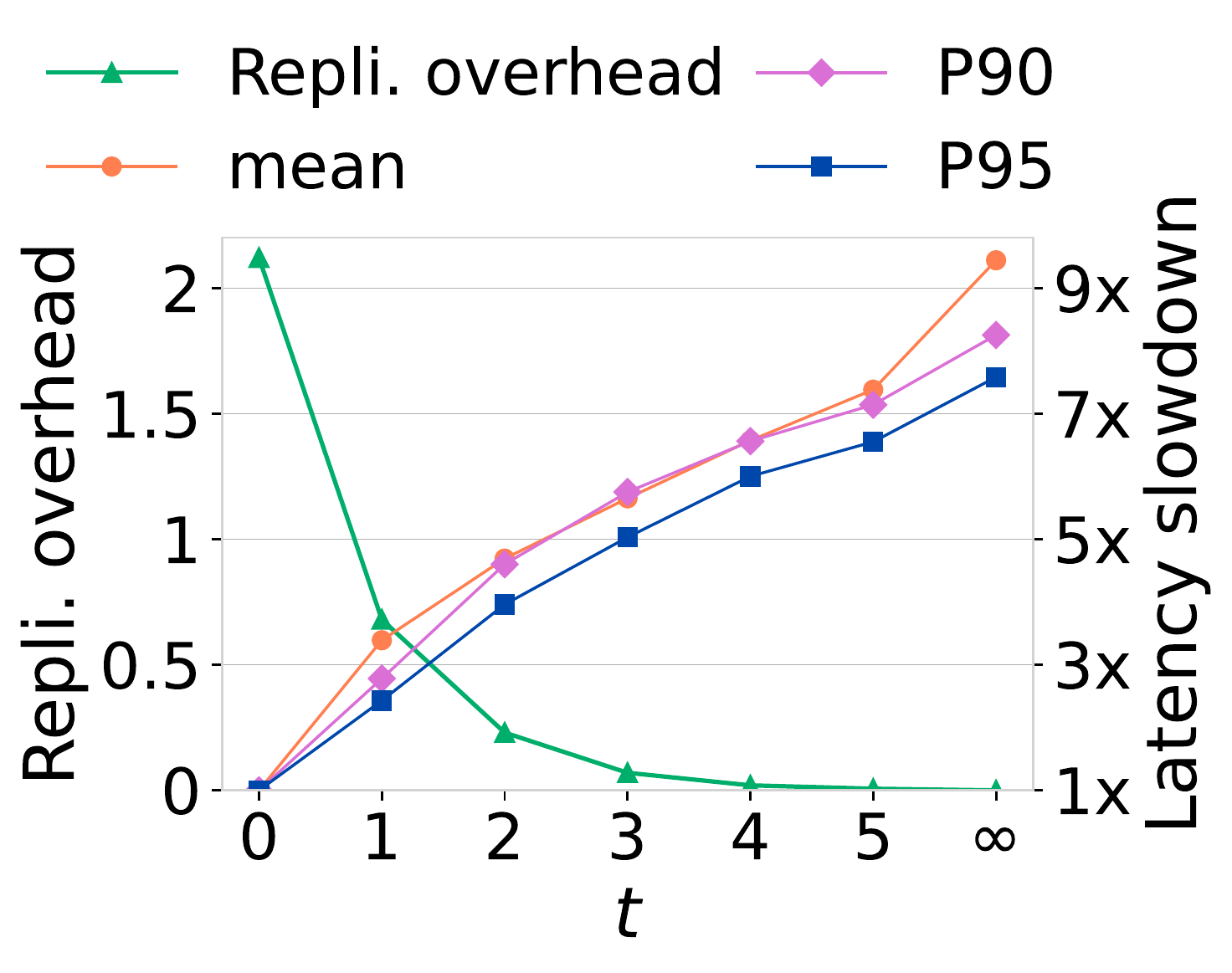}
    \caption{SNB: relative lat.\\ vs. replication}
    \label{fig:short_q_latency_normalized}
  \end{subfigure}\hspace{0.1mm}
  \begin{subfigure}[b]{.161\linewidth}
    \centering
    \includegraphics[width=\textwidth]{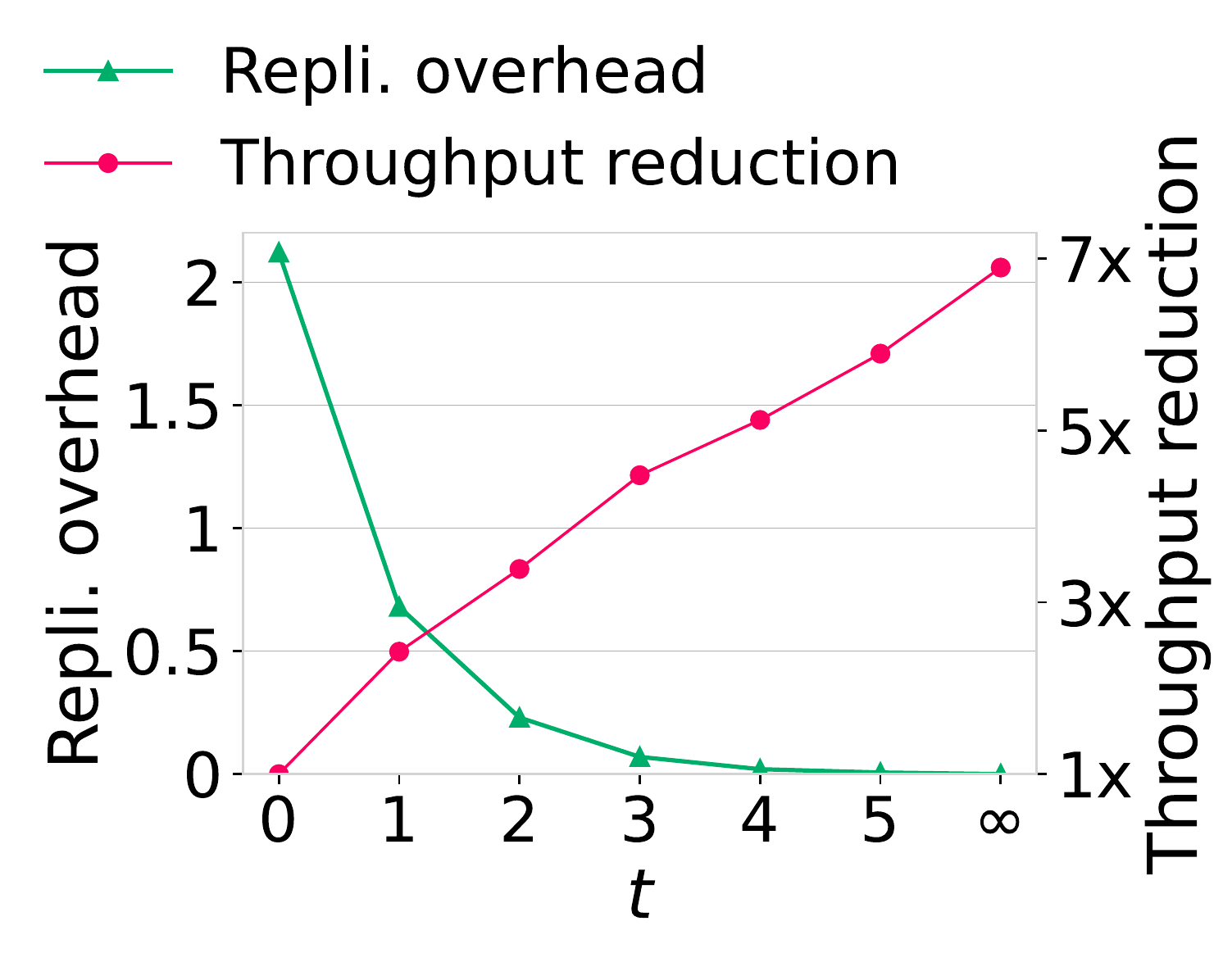}
    \caption{SNB: relative\\ tput vs. replication}
    \label{fig:short_q_throughput}
  \end{subfigure}
  \hspace{0.2mm}
  \begin{subfigure}[b]{.156\linewidth}
    \centering
    \includegraphics[width=\textwidth]{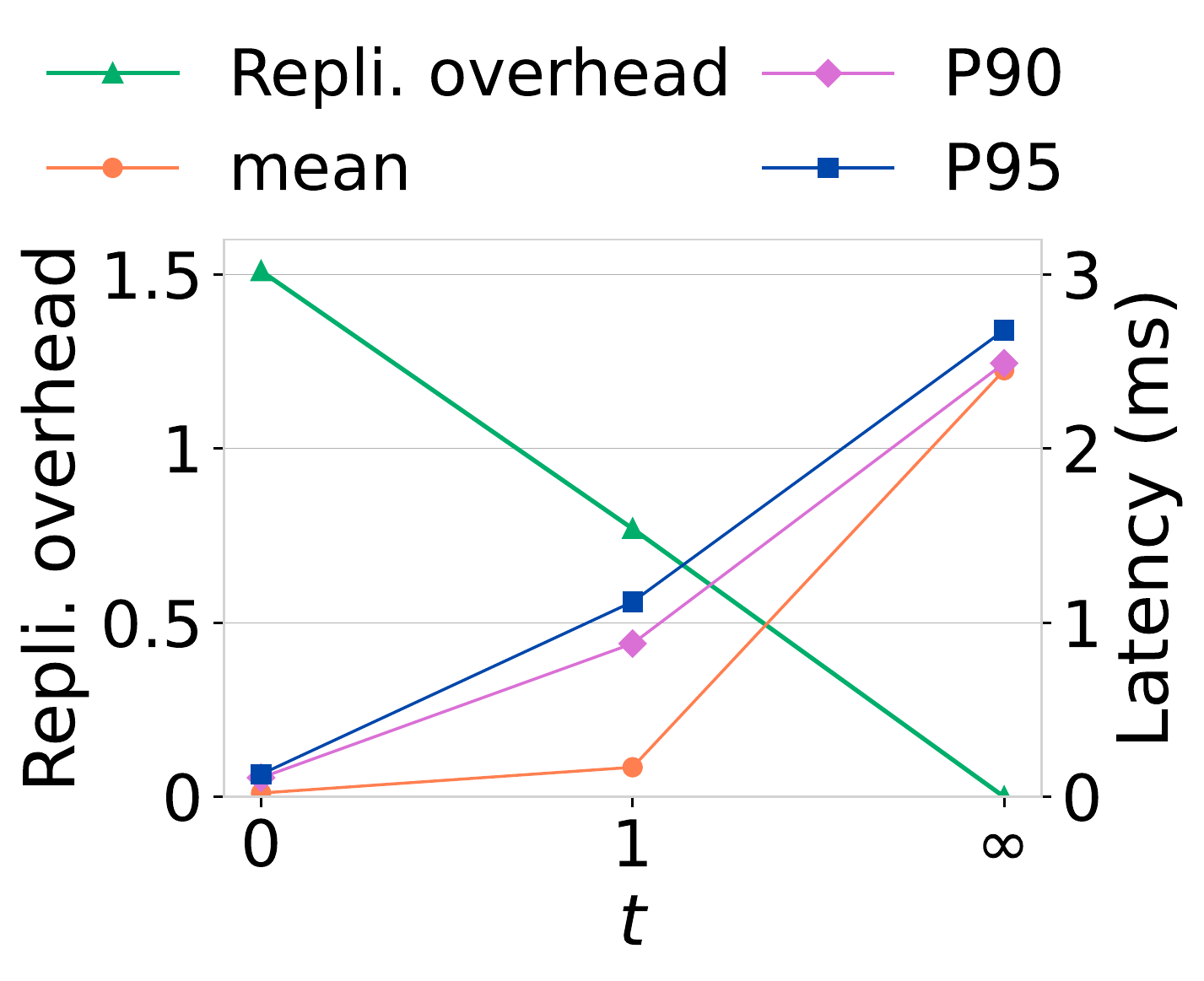}
    \caption{Sampling: latency\\ vs. replication}
    \label{fig:gnn_latency_not_normalized}
  \end{subfigure}\hspace{0.1mm}
  \begin{subfigure}[b]{.166\linewidth}
    \centering
    \includegraphics[width=\textwidth]{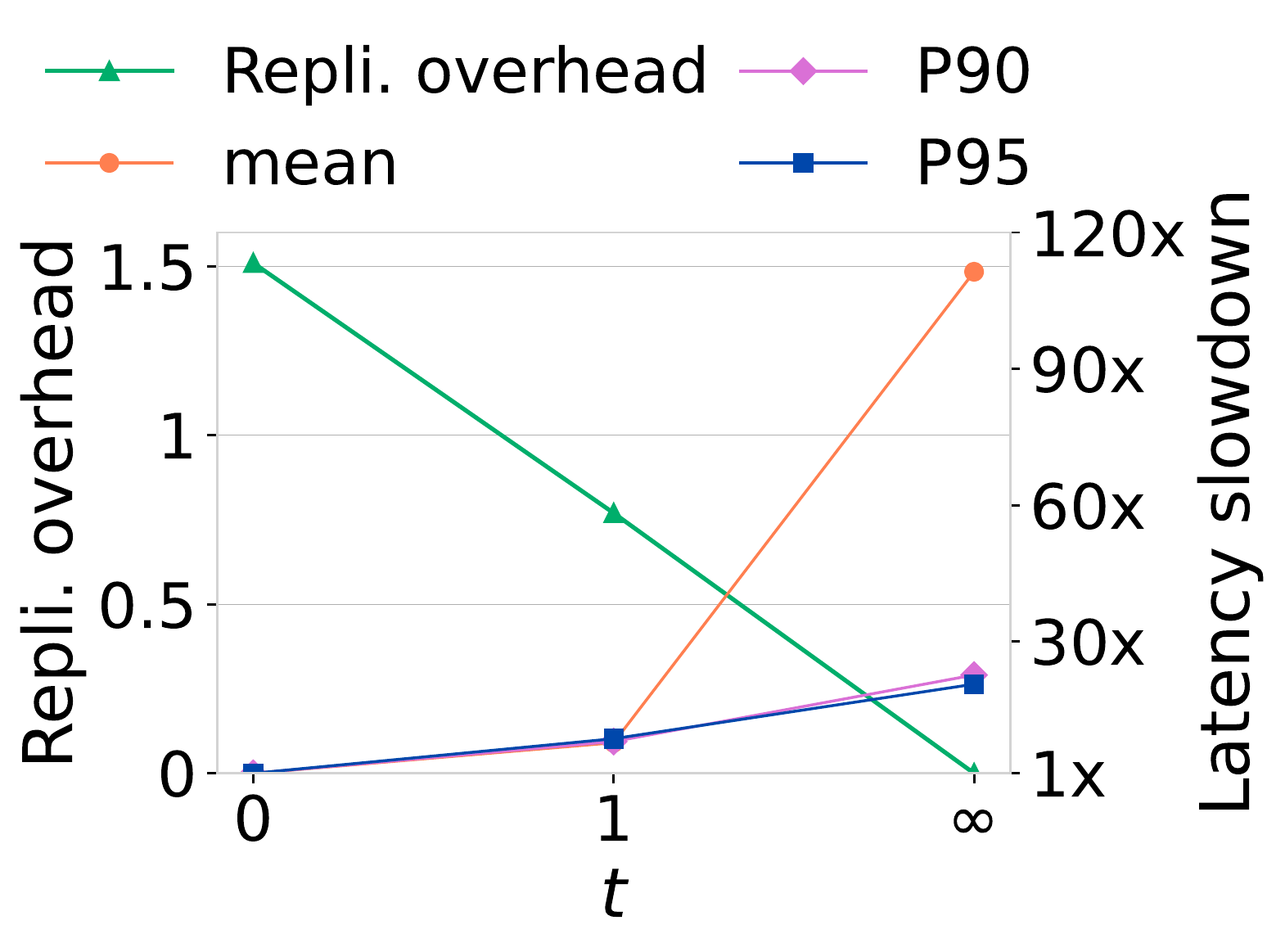}
    \caption{Sampling: relative lat. vs. replication}
    \label{fig:gnn_latency_normalized}
  \end{subfigure}\hspace{0.1mm}
  \begin{subfigure}[b]{.168\linewidth}
    \centering
    \includegraphics[width=\textwidth]{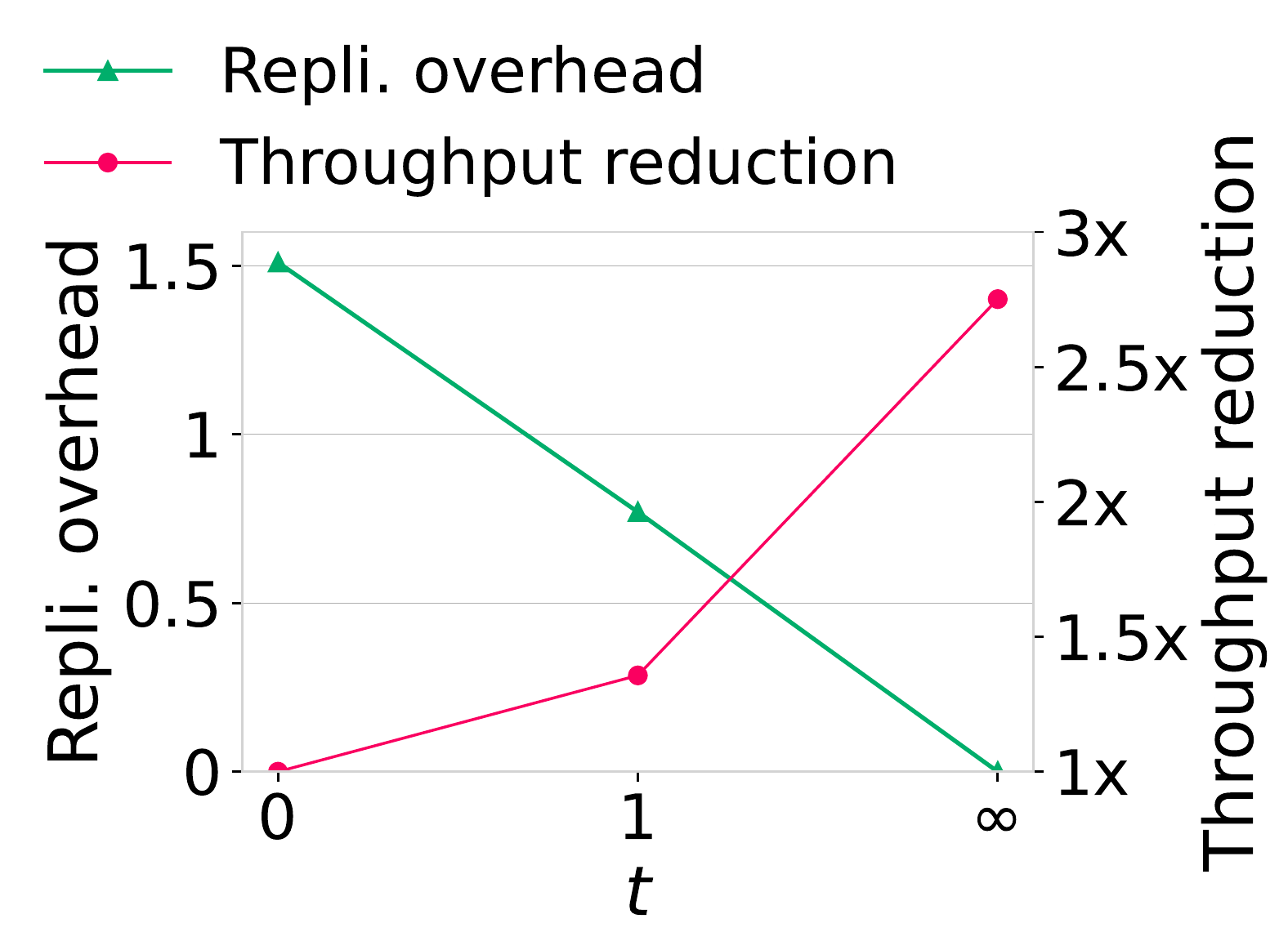}
    \caption{Sampling: relative\\ tput vs. replication}
    \label{fig:gnn_throughput}
  \end{subfigure}
  \vspace{-0.5\baselineskip}
  \caption{Query latencies and system throughput with varying $t$}
  \label{fig:q_latency}
  \vspace{-0.5em}
\end{figure*}

Our evaluation aims to answer the following questions:
\begin{itemize}
	\item[\textbf{Q1}] Can we effectively tune the tail latency of queries using replication and find better replication cost tradeoffs than existing approaches? 
	\item[\textbf{Q2}] Is our replication algorithm fast enough to be practical?
	\item[\textbf{Q3}] Can we tune throughput similar to latency?
	\item[\textbf{Q4}] Do the previous results generalize to a diverse set of data sharding schemes?
	% \item[\textbf{Q5}] Is it feasible to incrementally update the replication scheme at low replication cost when the dataset is updated?
\end{itemize}
%We first discuss the experimental setup and then answer all these questions in the positive.

\subsection{Experimental Setup}
\sparalight{Hardware configuration} 
%Our evaluation goal to study the latency-replication tradeoff using real-world workloads. 
We run the benchmarks on six AWS EC2 r5d.4xlarge instances, each with 16 virtual CPUs, 128GB RAM, 10 Gbps network, and Ubuntu Server 20.04 LTS (HVM). 
We run our replication algorithm on a server with two 16-core Intel Xeon(R) Silver 4216 2.1GHz processors, 128GB RAM, and Ubuntu 18.04. 

%\spara{Distributed system} We implemented a distributed query execution system to execute the benchmarks as described in Section~\ref{sec:system-model}.
%%The implementation is in C++ and use Apache Thrift \cite{thrift} as RPC framework. 
%%Clients initially send each query to one server.
%%When a server receives a query (or sub-query), it processes it by fetching data from local graph storage. 
%%If a required vertex is not stored locally, the server sends a sub-query to a remote server, selected using thrifty routing, using an RPC. 
%%The latency constraint $t$ on the number of distributed traversals is a bound on the maximum depth of nested RPCs required to complete the query.
%To store the graph, we use the LiveGraph storage system \cite{livegraph,lgrepo} for SNB and the Compressed Sparse Row (CSR) format for GNN sampling.
%%After all hops of the query are matched, the server returns the query result to the client. 

\sparalight{Benchmarks}
We evaluated our approach using two diverse benchmarks: the LDBC’s Social Network Benchmark (SNB) interactive short-read workload \cite{ldbc} and neighborhood sampling on the Open Graph Benchmark (OGB) \cite{OGB}. 
Table \ref{tab:data_graph} lists the statistics of the data graphs, considering different scale factors for SNB. 
The default data graphs are SNB  SF30 and OBG-papers100M.
All queries have the same latency constraint $t$, which is a parameter of our evaluation.

For SNB, we use hash partitioning as the default sharding scheme as it is common in distributed graph databases. 
The SNB query execution system we use is based on the open-source SNB interactive implementation on top of the LiveGraph storage system~\cite{livegraph-repo,livegraph-snb}.
This implementation outperforms existing commercial graph databases such as TigerGraph by more than one order of magnitude in our target benchmark (see~\cite{livegraph}) and is thus best suited to test the limits of low-latency query execution.
We extended that single-server implementation to run on a sharded distributed system.

\begin{comment}
The SNB workload consists of query types implemented as functions with input parameters, but each function accesses vertices and edges whose labels are statically determined.
For example, the query type of Figure~\ref{fig:causal_access_tree} accesses vertices and edges with statically-defined labels (\emph{person}, \emph{knows}, or \emph{creates}), while the root vertex (\emph{Alice}) is an input parameter.
Our workload analyzer considers all possible paths based solely on the label information.
Any data access that is conditional to the value of some parameter is assumed to be always executed.
This overapproximates the set of causal access trees that actually arise in the workload but it makes it faster to compute a replication scheme since it makes it unnecessary to enumerate all possible combinations of parameter values.
The analyzer aggressively prunes the data graph by processing one query type at a time, considering only edges and vertices having the labels required by the query type, and discarding properties that are not required to determine potential data accesses.
\end{comment}

OGB is a well-known benchmark for machine learning on graphs.
We use its \emph{ogb-papers100M} and \emph{ogb-mag} datasets. 
Building mini-batches for Graph Neural Network (GNN) training using Stochastic Gradient Descent (SGD) requires sampling the input graph.
Sampling a mini-batch is required before each training iteration and is the bottleneck when the graph is distributed~\cite{DistDGL}. 
We consider the common node-wise neighbor sampling method, which is used, for example, by GraphSAGE~\cite{graphsage}.
Specifically, we sample vertices at a distance up to three from the root, using a fan out of $25$ neighbors in the first hop and $10$ in the other two hops.
%Other than hash partitioning, we also use the Metis graph partitioner~\cite{METIS} to obtain the initial data placement scheme, as done by DistDGL~\cite{DistDGL}.
Our evaluation used a distributed sampling system akin to DistDGL~\cite{DistDGL}.
It stores adjacency lists in CSR format and uses the Metis graph partitioner by default~\cite{METIS}.
Sampling queries require no more than $2$ hops since the vertices in the $3^{rd}$-hop can be sampled from the adjacency list of the $2^{nd}$-hop vertex.
%The workload analyzer enumerates the $2$-hop paths in the graph and prunes the graph features.

\sparalight{Replication algorithm} We implemented the replication algorithm as a lock-free parallel algorithm in Java.
Each thread is assigned an equal-sized set of starting vertices of causal access paths in the workload. 
We used 64 threads in our evaluation.
Replicating a vertex on a server is represented by flipping a bit in a bit vector from $0$ to $1$.
Since replicas are never removed, updates from $1$ to $0$ never occur, so concurrent updates do not require acquiring locks.
% Compared to a serial implementation, this concurrent implementation may introduce extra replication costs as threads working on different paths may not see each others' updates.
% In practice, the difference was never more than $1\%$ in our experiments. 
We use a load imbalance constraint $\epsilon$ of 2\%. 
Our algorithm returned a feasible replication scheme for all the configurations we considered.

\begin{table}[t!]
\begin{small}
  \setlength{\tabcolsep}{12pt}
  \begin{tabular}{c|cc}
  \hline
  \multicolumn{1}{c|}{\textbf{Data graph}} &
    \multicolumn{1}{|c}{\textbf{Vertices}} &
    \multicolumn{1}{c}{\textbf{Edges}} \\ \hline
  SNB SF1  & 3,181,724  & 20,110,868 \\ 
  SNB SF3  & 9,281,922  & 62,665,752 \\ 
  SNB SF10 & 29,987,835 & 213,490,074 \\ 
  SNB SF30 & 88,789,833 & 664,628,912 \\ 
  OGB-mag & 1,939,743 & 21,111,007 \\ 
  OGB-papers100M & 111,059,956 & 1,615,685,872 \\ \hline
  \end{tabular}
\end{small}
  \caption{Data graph statistics}
\label{tab:data_graph}
\end{table}

\subsection{Evaluation Results}

\spara{Fine-Tuning Latency vs. Replication (Q1)}
We empirically show that the proposed replication algorithm enables users to fine-tune the query latency by controlling the latency constraint $t$.
Fine-tuning latency constraint enables finding sweet spots between latency and replication in both workloads that cannot be found using existing approaches.
In this evaluation, we assume that the upper bound $t$ is the same for all queries in the workload and we report aggregate latency across all queries in the workload.

\sparalight{Baselines}
As mentioned in Section~\ref{sec:background}, this is the first work that supports arbitrary query latency bounds, so we do not have replication algorithms to directly compare to.
A possible simplification of our work would be to require single-site execution, which is equivalent to only supporting a latency constraint of $t=0$.
We show by supporting arbitrary latency bounds, we significantly reduce the replication cost.

Some systems replicate the immediate remote neighbors of vertices to remove dangling edges between servers \cite{Wukong,DistDGL}.  
This could be seen as a replication scheme that enforces a specific latency constraint $t = n-1$, where $n$ is the maximum height of the causal access paths of the query.
If the adjacency list of the neighboring vertices is also replicated, a constraint of $t = \lfloor n/2 \rfloor$ can be enforced.
Such query-dependent latency bounds may not be practically useful in general, and that is why this optimization is usually introduced to improve average latency, not to bound tail latency.
However, we can still consider this as a baseline and compare to it, since this is an optimization that is also used in DistDGL~\cite{DistDGL}.
We consider the variant where we replicate also the adjacency list of neighboring vertices and show that our algorithm can enforce the same latency constraint with lower replication costs.

\sparalight{SNB benchmark}
Figure \ref{fig:short_q_latency_not_normalized} shows the mean and tail latency of SNB queries with varying $t$ and the replication cost required to provide such latency guarantee using the proposed replication algorithm. 
The notation $t = \infty$ means that we do not provide any latency constraint, and the system is in the original configuration without data replication.
The replication overhead is measured as the size of the additional replicated data over the size of the original dataset (i.e., the data graph).
In Figure \ref{fig:short_q_latency_normalized}, we normalize the query latency by the latency of the single-site execution (i.e., $t = 0$) to study the relative slowdown. 
Both mean and tail latency increase linearly with the latency constraint $t$.
This shows that the proposed replication algorithm can fine-tune latency by controlling $t$ as our latency model accurately reflects the actual latency of queries. 
These results show that we can tune latency in the sub-millisecond to single-digit millisecond ranges required by production application such as knowledge graphs~\cite{A1}.

As expected, requiring single-site execution and setting $t=0$ entails a high replication cost where we need to store three times as much data.
The replication overhead drops superlinearly as we linearly relax the latency constraint, following the trend shown in Figure~\ref{fig:tradeoff}, and this makes it possible to find a sweet spot balancing latency and replication. 
For instance, at $t=3$, further loosening the latency constraint does not save much replication, while tightening it becomes increasingly expensive.

\begin{figure*}[t!]
  \centering
  \begin{subfigure}[b]{.22\linewidth}
    \centering
  \includegraphics[width=\textwidth]{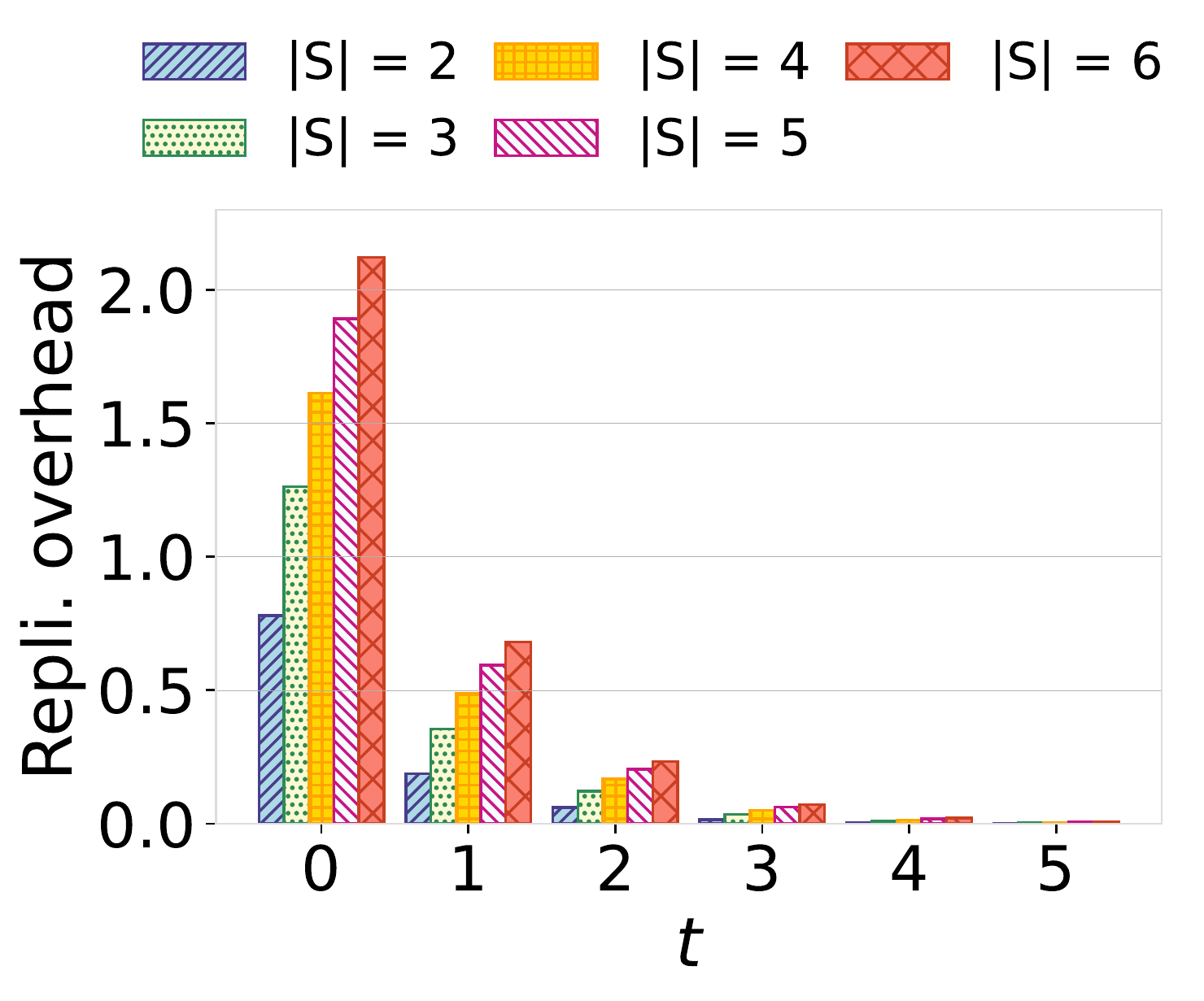}
  \caption{Hash\\ partitioning}
  \label{fig:short_q_replication_cost_hash}
  \end{subfigure}\hspace{1mm}
\begin{subfigure}[b]{.22\linewidth}
  \centering
  \includegraphics[width=\textwidth]{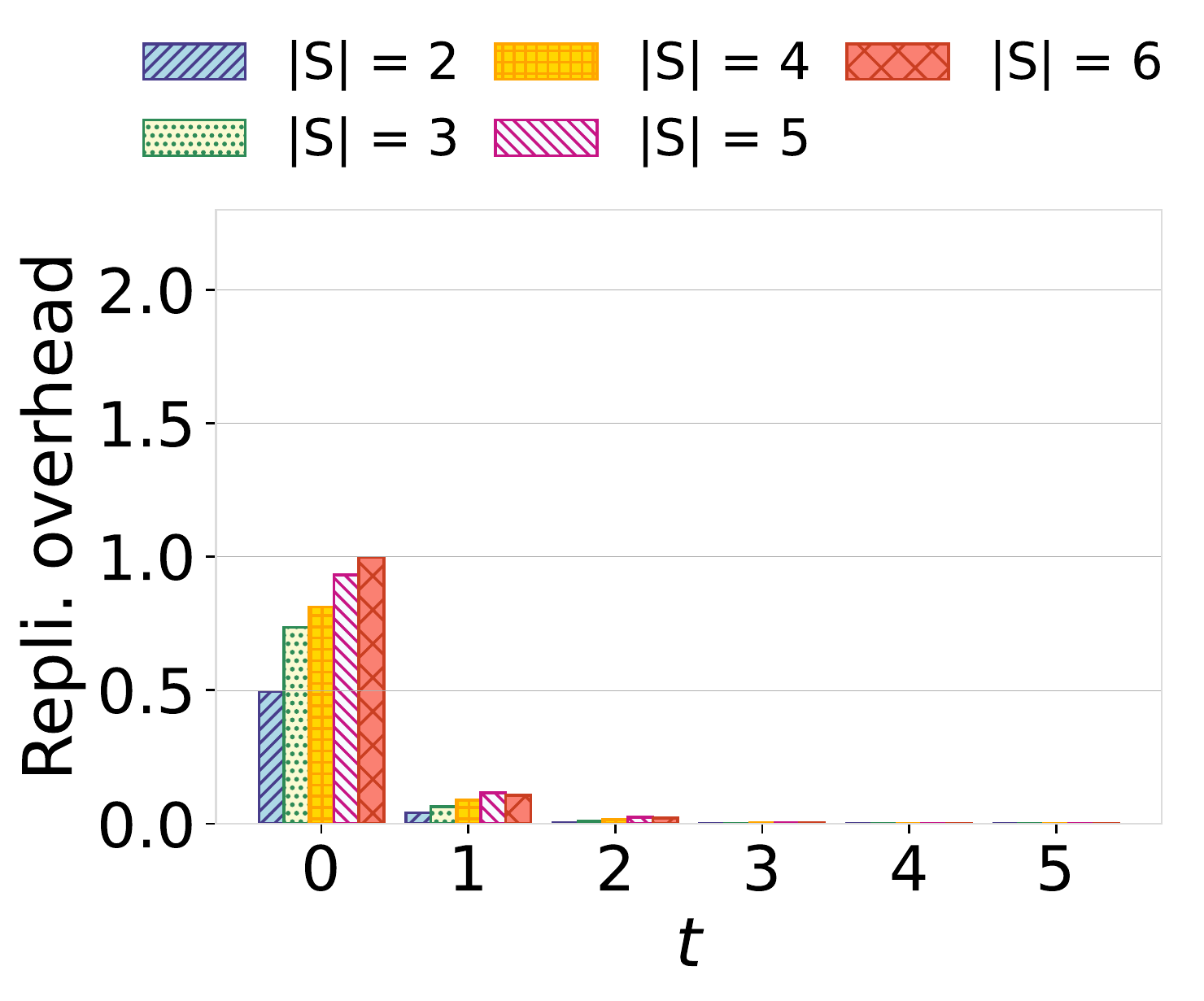}
  \caption{Graph\\ partitioning}
  \label{fig:short_q_replication_cost_metis}
  \end{subfigure}\hspace{1mm}
\begin{subfigure}[b]{.22\linewidth}
  \centering
  \includegraphics[width=\textwidth]{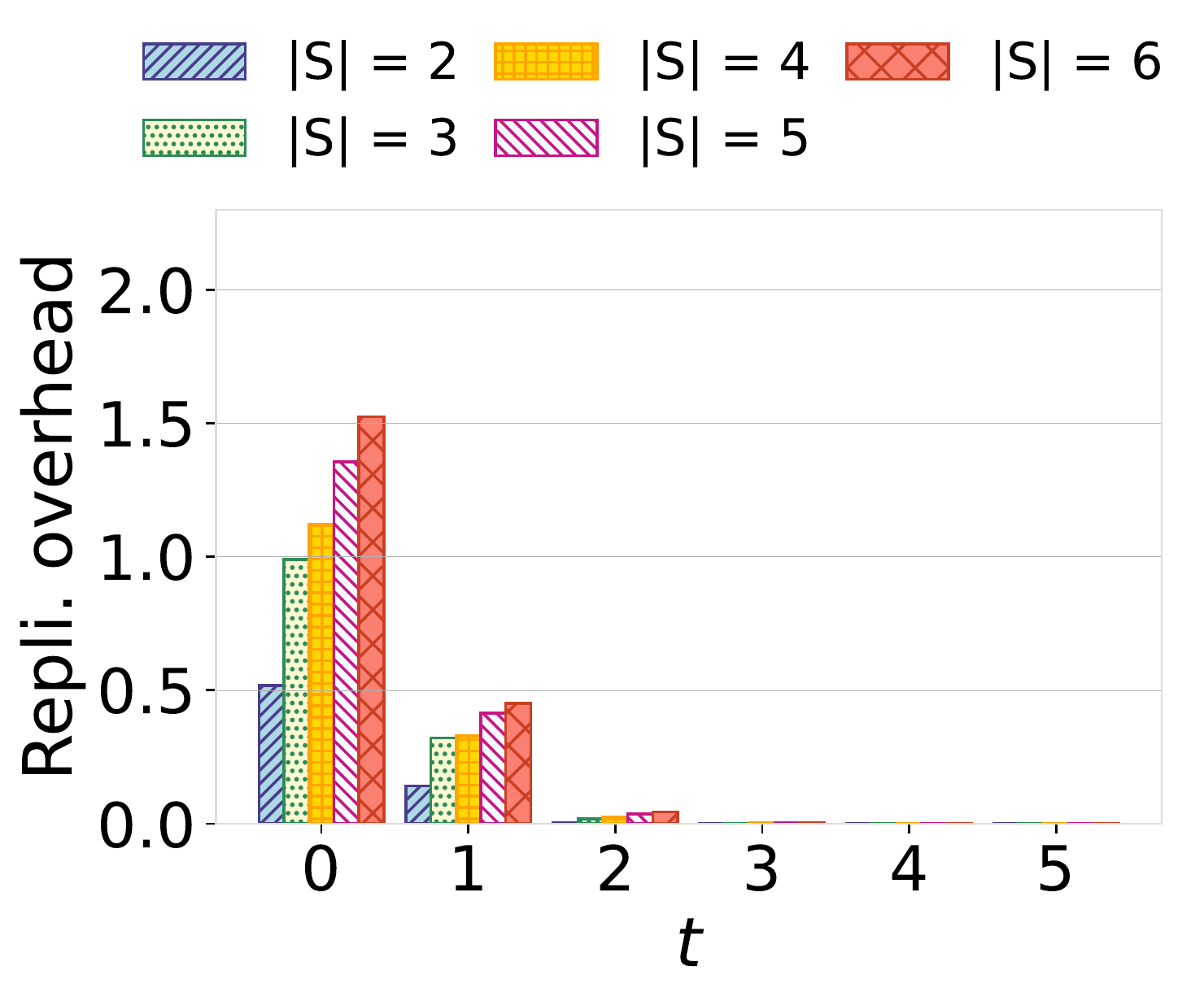}
  \caption{Hypergraph\\ partitioning}
  \label{fig:short_q_replication_cost_hmetis}
\end{subfigure}\hspace{1mm}
\begin{subfigure}[b]{.22\linewidth}
  \centering
  \includegraphics[width=\textwidth]{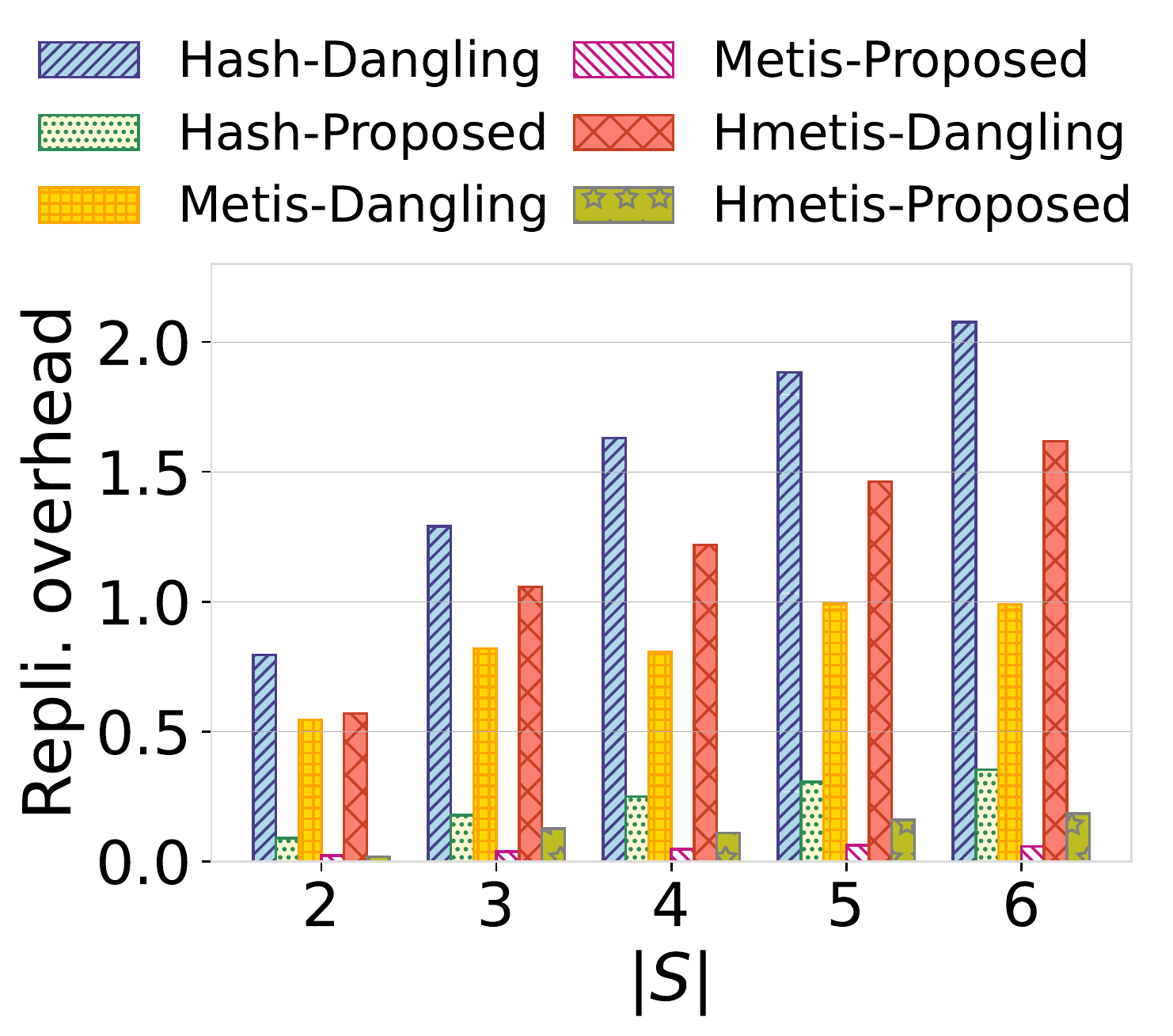}
  \caption{Proposed vs remove\\ dangling edges}
  \label{fig:proposed_vs_dangling}
\end{subfigure}
% \begin{subfigure}[b]{.192\linewidth}
%   \centering
% \includegraphics[width=\textwidth]{figures/exp/short_queries_dynamic_cost.pdf}
%   \caption{varying dynamic edge \%\\ ($|S|$ = 6)}
%   \label{fig:short_q_dynamic_cost}
% \end{subfigure}
\vspace{-0.5\baselineskip}
\caption{SNB workload replication cost}
\label{fig:short_q_replication_cost}
\vspace{-0.5em}
\end{figure*}

The comparison with dangling edges replication is reported in Figure~\ref{fig:proposed_vs_dangling}.
The proposed algorithm can achieve a much lower replication cost across all initial data placement schemes. 
This is because it is workload-aware and only replicates data based on the actual workload access patterns, whereas dangling edges replication only relies on the graph structure to decide what to replicate.

\sparalight{GNN sampling}
For GNN sampling, tail latency grows linearly as in the SNB case (see Figures~\ref{fig:gnn_latency_not_normalized} and~\ref{fig:gnn_latency_normalized}).
The relative slowdown compared to single-site execution ($t=0$) is larger since sampling queries are less computationally expensive than SNB queries.
With $t=1$, the mean latency grows less steeply because, on average, queries are still more likely to be fully local thanks to Metis-based sharding.
It is however likely that some queries will access remote vertices, which explains why tail latency grows linearly.
Without enforcing any latency constraint ($t=\infty$), most queries perform two distributed traversals, so mean and tail latencies are similar.

The replication overhead decreases almost linearly in this case since the likelihood of accessing any neighbor of a vertex is uniform, unlike in SNB where accesses are based on the label of the vertices and edges.
Setting $t=0$ results in high replication overhead also in this workload.
Setting $t=1$ represents the sweet spot because it substantially decreases the replication cost, keeps tail latency capped, and only increases the mean latency minimally.
Compared to dangling edges replication, we observe a similar trend in the GNN sampling workload as in SNB (see Table~\ref{tab:gnn_repli_cost}).

\spara{Running Time of the Algorithm (Q2)}
The running time of the workload analysis and replication algorithm are short enough to make it practical for a one-off offline analysis.
We report the minimum and maximum running time with varying $t$ and $|S|$ in Table \ref{tab:algo_running_time}. 
The running times include the time required to generate causal access paths using the workload analyzer, which happens on-the-fly as the replication algorithm is executed.
For SNB, the replication time increases linearly with the graph size, and the replication algorithm can finish in less than three minutes for the largest SNB data graph we tested, which has more than 600 million edges.  
For the GNN sampling workload, the replication algorithm finishes in at most 22 minutes for the largest graph on OGB with one billion edges. 
Without the redundant path pruning optimization of Section~\ref{sec:rep-algo}, the running time of the algorithm exceeds one hour in all cases except SNB SF1.

\spara{Fine-Tuning Throughput (Q3)}
We observe that tuning latency constraints is also an effective way to control throughput.
Figures~\ref{fig:short_q_throughput} and~\ref{fig:gnn_throughput} report the relative throughput with varying latency constraint $t$ for the two workloads.
The throughput with $t=0$ is about 80k and 15k queries per second for the SNB and sampling workload, respectively.
The trends for throughput are similar to the ones for latency.

\spara{Varying Sharding Schemes (Q4)}
We now evaluate the impact of different initial data placement schemes on the replication cost.
We consider three sharding schemes: \emph{hash partitioning}, \emph{graph partitioning}, and \emph{hypergraph partitioning}. 
Hash partitioning is common in practice because of its simplicity, and it does not require preprocessing. 
Graph partitioning is data-aware and workload-unaware since it only considers the data graph for partitioning.
We use Metis~\cite{METIS} to partition the data graph. 
Hypergraph partitioning is both data- and workload-aware and is inspired by \cite{schism,social_hash}. 
We sample and model the workload as a hypergraph, where a data vertex is represented as a node, and all data vertices accessed in a query are grouped into a hyperedge. 
We run 1M queries to build the hypergraph and partition it using hmetis~\cite{hmetis}. 
None of these algorithms can enforce latency constraints. 
We use them as initial data placement schemes and run our algorithm to provide latency constraints.
We focus our evaluations on the SNB workload.

Requiring single-site execution and setting $t=0$ results in high replication overhead across all initial placements schemes, especially when the system has more servers.
With $6$ servers, the additional replication overhead on top of the initial storage cost is larger than 2 across all schemes.
In general, more servers typically result in more queries requiring distributed traversals, which require more replicas to optimize. 
The replication overhead decreases with the latency bound. 

Hash partitioning (Figure \ref{fig:short_q_replication_cost_hash}) results in the highest overhead because it randomly assigns vertices to servers without considering the workload or the data graph structure. 
Graph partitioning (Figure \ref{fig:short_q_replication_cost_metis}) requires the least replication cost.
Hypergraph partitioning (Figure \ref{fig:short_q_replication_cost_hmetis}) results in higher replication cost even though it is workload-aware.
This is because workload traces only capture incomplete information of the entire data graph.

\begin{table}[]
\begin{small}
  \begin{tabular}{c|ccc}
  \hline
  \textbf{Dataset} & \textit{\textbf{k = 0}} & \textit{\textbf{k = 1}} & \textit{\textbf{No Dangling Edges}} \\ \hline
  OGB-mag (Hash)   & 1.51  & 0.77 & 1.48 \\ 
  OGB-mag (Metis)  & 0.48  & 0.06 & 0.23 \\ 
  OGB-papers100M (Hash)  & 2.69  & 1.48 & 2.33 \\ \hline
  \end{tabular}
  \end{small}
  \caption{Replication overhead on the GNN sampling workload with different initial data placement schemes ($|S|$ = 6). }
  \label{tab:gnn_repli_cost}
  \end{table}

\begin{table}[]
\begin{small}
  \begin{tabular}{cc|cc}
  \hline
  \textbf{Data graph}     & \textbf{Workload} & \textbf{Min time (s)} & \textbf{Max time (s)} \\ \hline
  SNB SF1  & SNB      & 1.2          & 3.7          \\ 
  SNB SF3  & SNB      & 8.6          & 24           \\ 
  SNB SF10 & SNB      & 25           & 65           \\ 
  SNB SF30 & SNB      & 118          & 173          \\ 
  OGB-mag & GNN       & 17           & 30           \\ 
  OGB-papers100M & GNN & 414          & 1290         \\ \hline
  \end{tabular}
  \end{small}
  \caption{Min \& max time for generating replication schemes. }
  \label{tab:algo_running_time}
\end{table}

\section{Related Work}
\label{sec:related}

\spara{Maximizing aggregate data access locality}
Prior work has explored data placement, replication, and migration techniques to maximize \emph{aggregate} data access locality across an \emph{entire workload}~\cite{MorphoSys,serafini2014accordion,taft2014store,Clay,Pragh,social_hash,hermes,squall,morphus,schism,sword,JECB,graph_partitioning,Clay,Wukong,DistDGL,managing_large_dynamic_graphs_efficiently,accelerate_SPARQL}.
This paper addresses for the first time a more challenging problem: ensuring an \emph{upper bound} on tail latency.
This requires different models and replication algorithms.

Some of this prior work~\cite{schism,sword,JECB,graph_partitioning,Clay} uses workload graphs to model the workload, which differ from causal access paths as discussed in Section~\ref{sec:models}.
Other work maximizes aggregate data access locality using different replication techniques~ \cite{Wukong,DistDGL,managing_large_dynamic_graphs_efficiently,accelerate_SPARQL}
or online data migration~\cite{MorphoSys,serafini2014accordion,taft2014store,Clay,Pragh,social_hash,hermes,squall,morphus}. 
%Some of the online data migration techniques introduced in previous work, such as Squall~\cite{squall}, could be used by query execution systems using our approach to update the replication scheme online.
%We did not consider using online data migration in our first solution to our new problem, but this represents an interesting future direction.

%MorphoSys \cite{MorphoSys} performs data migration and re-mastering before executing the query to ensure single-site transactions.
%Clay \cite{Clay} periodically builds a workload monitoring graph to balance load while minimizing distributed transactions.
%Pragh \cite{Pragh} proposes a fine-grained migration approach to preserve locality for RDMA systems. 
%Social Hash \cite{social_hash} is a data partitioning framework that optimizes the operations of large social networks.
%Hermes \cite{hermes} proposes a dynamic repartitioning algorithm to maintain partition quality online. 
%Squall \cite{squall} and Morphus \cite{morphus} support online repartitioning of databases using data migration. 

\spara{Enforcing single-site execution}
Our work differs from work enforcing single-site execution by allowing to fine-tune the latency bound and relaxing some assumptions on the workload while focusing on the latency of read queries, as discussed in Section~\ref{sec:background}.
DynaMast dynamically remasters objects to provide single-site execution and fully replicates the database at all servers~\cite{abebe2020dynamast}.

\spara{Sharding systems}
Sharding frameworks allow abstracting away the sharding, replication, and request routing logic of an application~\cite{slicer,kakivaya2018service,lee2021shard}.
%They support application-defined data placement policies to achieve data access locality, grouping frequently co-accessed data items into a shard, and store shards across servers.
Compared to monolithic implementations, integrating query execution systems with these frameworks could make it easier to integrate data placement and replication algorithms like ours in practical production systems.

\spara{Optimizing distributed graph access}
Much prior work on graph databases aims to improve the performance of distributed graph queries~\cite{grasper,gcache,gtran,graphrex,trinity,trinity_rdf}.
It does not aim to bound tail latency, as our work does.
Making distributed traversals faster can allow users to relax their latency bounds and thus to relax the replication cost of enforcing those bounds.
%Chiller \cite{chiller} proposes a new approach to perform data partitioning and distributed transactions to minimize data contention. 
%Grasper \cite{grasper} proposes a query execution model for OLAP on property graphs.  
%\cite{fast} proposes a multi-way join approach for distributed complex joins.
%GCache \cite{gcache} proposes a caching mechanism to speed up graph algorithms by utilizing the topological properties of the data graph. 
%G-Tran \cite{gtran} is an RDMA-based graph database that optimizes graph transactions. 
%\cite{graphrex} proposes a framework for graph processing on data center infrastructure. 
%Trinity \cite{trinity} is a general-purpose graph engine over a distributed memory cloud that optimizes memory management and network communications. 
%Trinity.RDF \cite{trinity_rdf} is a distributed, memory-based graph engine for RDF data that supports efficient processing of SPARQL queries and graph algorithms. 
%All the aforementioned work are \emph{best-effort} approaches and do not provide latency constraints on queries. 

\section{Conclusion}
This work shows for the first time that it is feasible to control the query tail latency in a workload but using replication and appropriate workload analysis and replication algorithms.
It lays the foundations of a general algorithmic framework that can be extended to other workloads, beyond the low-latency graph queries targeted by this work.
One interesting future extension is developing query execution systems that intertwine incremental updates to the replication scheme with the processing of write queries.
Another possible extension is to consider other types of bottlenecks than distributed traversals.
Finally, extending this work to relational workloads is an interesting potential research avenue.

%enable users to flexibly set the latency constraints of multi-hop queries. 
%Our evaluation confirms that our algorithms enable tuning the query latency by bounding the number of distributed traversals it requires. 
%One future direction is to explore how to adopt this technique in different settings, such as optimizing for micro-services on cloud. 
%Another is how to use data replication to balance query locality and execution parallelism instead of simply prioritizing locality. \hl{Should we say something about the future work for dynamic updates?}

%%
%% The acknowledgments section is defined using the "acks" environment
%% (and NOT an unnumbered section). This ensures the proper
%% identification of the section in the article metadata, and the
%% consistent spelling of the heading.
% \begin{acks}
% To Robert, for the bagels and explaining CMYK and color spaces.
% \end{acks}

%%
%% The next two lines define the bibliography style to be used, and
%% the bibliography file.
\bibliographystyle{plain}
\bibliography{references}

\newpage

\newpage
\begin{appendices}
\section{Proofs}
This appendix contains the proofs omitted from the main paper because of lack of space.

\subsection{Proof of Theorem~\ref{thm:NP-hardness}}
We show that checking that there exists a replication scheme satisfying a given latency bound and storage capacity constraint for each server, called \emph{the latency-storage feasible problem}, is NP-hard. This implies NP-hardness of the latency-bound replication problem.   The proof has two steps. Our key insight is in the first step, where we introduce the \emph{min-bridge bisection problem}: given a graph, the task is to partition the graph into two subgraphs with an equal number of vertices such that the number of vertices with neighbors in the other partition is bounded. We then reduce the latency-storage feasible problem to the min-bridge bisection problem. 
	%In the first step, we reduce the latency-storage feasible problem to a problem we call \emph{min-bridge bisection problem}, where we want to partition the graph into two subgraphs with an equal number of vertices such that the number of vertices with neighbors in the other partition is bounded.
In the second step, we reduce that problem to the min-bisection problem for 3-regular graphs, which is NP-hard~\cite{bui1987graph}. The two reductions imply that the latency-storage feasible problem is NP-hard.

\sparalight{Step 1.}
The (decision version of the) min-bridge bisection problem is defined as follows: given a graph $G = (V_G, E_G)$ with $2n$ vertices and a parameter $K$, check if there exists a  partition of $V_G$ into two sets $V_1,V_2$ of $n$ vertices each such that the number of ``bridge vertices'' in each set is at most $K$, where a bridge vertex is a vertex in one set having at least one neighbor in the other set.

Given a graph $G$ in the min-bridge bisection problem, we build an instance of the latency-bound replication problem as follows.  First, we build a dataset $D$ and a workload $W$.
For each vertex $v \in V_G$, we add two objects to $D$: a \emph{regular} object $v_o$ and a \emph{marker} object $v_m$.
We also add a query to $W$ with the following set of causal access path: in each path, $v_m$ is the root with a child $v_o$. Then for each path we make $v_o$ the parent of the regular object $u_o$ for each vertex $u \in N(v)$.  The storage cost function $f$ is such that $f(v_m) = 1$ for each marker object and $f(v_o) = 1/(2n)$ for each regular object. We set the balancing parameter $\epsilon = +\infty$ to disregard the load balancing constraint. 
 %for some $\epsilon < 1$, where $\epsilon < 1$ is the load balancing constraint of Eqn.~\ref{eqn:problem}. 
 We have four servers.     The maximum storage of the servers are $M_{s_1} = M_{s_2} = n+1/2$ and $M_{s_3} = M_{s_4} = n+ 1/2 + K/(2n)$. The sharding function $d$ is defined as follows. Servers $s_1$ and $s_2$ contain the marker objects for half of the vertices in $V_G$ each.
Server $s_1$ also contains the regular objects for the vertices whose markers are in $s_2$ and vice versa. 
The latency bound for all queries is $0$. We denote the instance of the latency-bound replication problem by $LS(G)$.

 We show that the  $G$ has a bisection with $K$ bridge vertices if and only if $LS(G)$ has a feasible solution.

For the if direction, we assume that $G$ has a bisection $(V_1,V_2)$ with at most $K$ bridge vertices in each set.    %Let $K_1$ and $K_2$ be the number bridge vertices in $V_1$ and $V_2$, respectively. Then $K_1 + K_2 = K$.  
We replicate the marker and regular objects of vertices in $V_1$, and the regular objects of the neighbors of  $V_1$ to server $s_3$. Similarly,  the marker and regular objects, and the regular objects of the neighbors of $V_2$ are replicated to $s_4$. Then, for every bridge vertex in $V_2$ ($V_1$), we replicate its \emph{regular object} in $s_3$ ($s_4$, respectively). Observe that $s_3$ has  storage cost $n+n\cdot 1/(2n) + K(1/(2n)) = n+1/2 + K/(2n)$. Similarly, $s_4$ has  storage cost $ n+1/2 + K/(2n)$. Thus, the latency and storage capacity constraints are satisfied. Observe that $s_1$ and $s_2$ satisfy the storage capacity constraint due to the way we set their capacity.  Hence $LS(G)$ has a feasible solution.

%As $K_1,K_2$ are both at most $n$, load balancing and maximum storage constraints are satisfied. Thus, the total storage cost of all 4 servers is $2(n+\epsilon) + 2(n+\epsilon) + (K_1 + K_2)(\epsilon/n) = 4n + 4\epsilon + K(\epsilon/n)$.

We now focus on the only if direction. We assume that $LS(G)$ has a feasible replication scheme $r$. Since the root of every causal access path is a marker object, without any replication, the marker of a vertex and its regular object are located on two different servers. Thus, accessing a marker at server $s_1$ or $s_2$ will result in a subsequent distributed traversal to the regular object, which would violate the latency bound. 
Therefore, $r$ must add replicas to servers $s_3$ and $s_4$ since $s_1$ and $s_2$ have already reached their storage capacity. %The load balancing constraint and maximum storage constraints of Eqn.~\ref{eqn:problem} requires that $s_3$ and $s_4$ have replicas for no more than half of the marker objects each plus replicas of any number of regular objects. 

%We now establish some properties about the solution of this instance of the latency-bound replication problem. The root of every causal access tree is a marker object.     Without replication, the marker of a vertex and its regular object are located on two different servers. Accessing a marker at server $s_1$ or $s_2$ will result in a subsequent distributed traversal to the regular object, which would violate the latency bound.   Therefore, we need to add replicas to servers $s_3$ and $s_4$ since $s_1$ and $s_2$ have already reached their maximum storage capacity. The load balancing constraint and maximum storage constraints of Eqn.~\ref{eqn:problem} requires that $s_3$ and $s_4$ have replicas for no more than half of the marker objects each plus replicas of any number of regular objects.

Since maximum storage capacity of $s_3$ and $s_4$ is $n+ 1/2 + K/(2n) < n+1$ as\footnote{if $K\geq n$, the problem is trivial; any partition of the vertices works.} $K\leq n-1$, $s_3$ and $s_4$ must contain exactly $n$ marker objects each. %If a server, say $s_3$,  contains exactly $n+1$ marker objects, it must contain additional $n+1$ regular objects corresponding to these marker objects by the latency constraint. Thus, the storage cost of $s_3$ is at least $(n+1)+ (n+1)(\epsilon /n) \geq n+1+\epsilon > n+2\epsilon$, violating the maximum storage cost bound. Thus, $s_3$ and $s_4$ contain exactly $n$ marker objects each.
Let $V_1$ and $V_2$ be the vertices whose marker objects are replicated to $s_3$ and $s_4$, respectively. We claimed above that $|V_1| = |V_2| = n$. Thus, $(V_1,V_2)$ is a bisection of $G$. Let $B_1$ and $B_2$ be the number of bridge vertices in $V_1$ and $V_2$, respectively. Let $K_1 = |B_1|$ and $K_2 = |B_2|$. Our goal is to show that $\max\{K_1,K_2\}\leq K$.  By the latency constraint, all regular objects of vertices in $V_1$ and $B_2$ must be replicated to $s_3$. Thus, the storage cost of $s_3$ is at least $n+n(1/(2n)) + K_2(1/(2n))= n+1/2 + K_2/(2n)$ which is at most the maximum storage of $s_3$ if and only if $K_2\leq K$. By the same argument, we get $K_1\leq K$ as desired.

\sparalight{Step 2.} In this step, we show that the min-bridge bisection problem is NP-hard by a reduction from the (decision version of the) min-bisection problem in 3-regular graphs, which asks for a bipartition of the vertices into two parts such that the number of edges crossing the bipartition is at most $K$. This problem is NP-hard~\cite{bui1987graph}. Given a 3-regular graph $G$ with $2n$ vertices, construct a graph $H$ by replacing each vertex $v$ by three copies $v_1,v_2,v_3$ that are connected to each other. Then for each edge $(u,v)$ in $G$ we add an edge from a degree-2 copy of $u$ to a degree-2 copy of $v$ in $H$. % , each attached to one of the edges of $v$ and having edges to the other two copies. 
The resulting graph $H$ has $6n$ vertices and is also 3-regular.

Now we claim that $G$ has a bisection of size $K$ if and only if $H$ has a bisection such that each set in the bisection has at most $K$ bridge vertices.

For the if direction, suppose that $G$ has a bisection of size $K$. Let the vertex partition of $G$ be $(V_1,V_2)$, each contains $n$ vertices. We then form the vertex partition $(U_1,U_2)$ of $H$ as follows: for each vertex $v$ in $V_1$, add all 3 copies to $U_1$. Since there are $K$ edges crossing $(V_1,V_2)$, there are $K$ bridge vertices in $U_1$ (and also in $U_2$), each corresponds to a crossing edge of $(V_1,V_2)$. %Thus, the total number of bridge vertices is $2K$.

For the only if direction, suppose that $H$ has a bisection $(U_1,U_2)$ with at most $K$ bridge vertices in each set.  First, we claim that all copies of every vertex $v$ in $G$ can be placed on one side of  $(U_1,U_2)$ without increasing the number of crossing edges.  This is because if the copies of the same vertex $v$ are on two sides, for example $v_1$, $v_2$ are in $U_1$ while $v_3$ in $U_2$, then there must be another vertex, say $w$, whose copies are on different sides of the partition (say $w_1$ in $U_1$ while $w_2$ and $w_3$ are in $U_2$).  We then swap $v_3$ to $U_1$ and $w_1$ to $U_2$. Observe that the number of crossing edges does not increase since $H$ is 3-regular. Furthermore, the resulting partition remains balanced. By keeping swapping copies, we finally get a partition $(U_1,U_2)$ where all copies of every vertex are on one side. 
%It’s not hard to see that the swapping operation does not increase the number of bridge vertices.

Now, as all copies of every vertex are on one side of $(U_1,U_2)$, by simply replacing these copies with the original vertex, we obtain a bipartition $(V_1,V_2)$ of $G$. Observe that each bridge vertex $v\in U_1$ has exactly one edge to another vertex in $U_2$ by the construction of $H$. Thus, the number of edges crossing $(U_1,U_2)$ is exactly $K$. This is also the number of edges crossing the bipartition of $G$.

\subsection{Proof of Theorem~\ref{thm:correct-update}}
To prove this Theorem, we first introduce the notion of extension of a replication scheme.

\begin{definition}[Extension of a replication scheme]
    Let $r$ be a replication scheme for a dataset $D$. 
    A replication scheme $r'$ is an extension of $r$ for some object $v' \in D$ if and only if:
    $$
        r(v') \subset r'(v') \wedge
        |r'(v') \setminus r(v')| = 1 \wedge
        \forall v \neq v' \in D, r'(v) = r(v)
    $$
\end{definition}

We have that:

\begin{lemma} \label{lem:inductive-robust}
    Let $r$ be a replication scheme that is latency-robust for $p$ and $r'$ be an extension of $r$. It follows that $r'$ is latency-robust for $p$.
\end{lemma}

This lemma directly follows from the fact that $r'$ only adds replicas to $r$, so it preserves the latency-robustness condition for $r$. 
% \begin{comment} == TO DISCUSS
% \hl{It seems to me that when the replication of one vertex could involve replicate more than one copies to different servers holding the master copy of the master objects of the ancestors in the path, so the condition $(|r'(v') \setminus r(v')| = 1)$ in the extension definition is not really accurate.}
% \ms{See my comment above.}\hl{I meant we should add $|r'(v') \setminus r(v')| \leq 1$ because $v'$ could be replicated to more than one servers under $r'$.}
% \ms{I see. I agree}
% \end{comment}
We can now prove the following result.

\begin{lemma}\label{thm:extend}
    Let $p$ be a root-to-leaf path in a causal access path of a query $Q$, $r$ a replication scheme, and $t_Q$ an upper bound on latency of $Q$. 
    If $r$ is robust for $p$ and $h(p,r,\rho) \leq t_Q$, for any replication scheme $r'$ extending $r$,  $h(p,r',\rho) \leq t_Q$. % there exists no replication scheme $r'$ extending $r$ such that $h(p,r',\rho) > t$. 
\end{lemma}
%\hl{$t$ should be $t_Q$. I found it more natural to say: If $r$ is latency-robust for $p$ and $h(p,r,\rho) \leq t_Q$, then for any extension $r'$ of $r$,  $h(p,r',\rho) \leq t_Q $.}
\begin{proof}
    Assume by contradiction that $r$ is robust for $p$ and there exists an $r'$ extending $r$ such that $h(p,r',\rho) > t_Q$.
    This implies that $h(p,r,\rho) < h(p,r',\rho)$, so $r'$ must change the access location of some object $v$ in $p$, i.e., $\rho(r,v) \neq \rho(r',v)$.
    %Let $a_M$ be the proposed routing function under $r$ and $a'_M$ be the proposed routing function under $r'$.
    %It must hold that $h(p, a_M) > 0$, since otherwise all vertices in $p$ can be matched at the server that contains the master copy of the root vertex of $p$, and there cannot exists an extension $r'$ of $r$ that causes a violation.
    %Therefore, matching $p$ requires at least one hop, that is, one inter-server RPC call.
    Let $p = \langle v_{p_1}, v_{p_2}, ..., v_{p_n}\rangle$, $m$ be the smallest index such that $\rho(r,v_{p_m}) \neq \rho(r',v_{p_m})$, and $g_{p,r}^i = \langle v_{p_j}, ..., v_{p_k}\rangle$ be the server-local traversal subpath under $r$ including $v_{p_m}$.
    By definition of server-local subpath, $\rho(r,v_{p_{j-1}}) \neq \rho(r,v_{p_j})$, so with $r$ the query accesses $v_{p_j}$ at server $s=d(v_{p_j})= \rho(r,v_{p_j})$.
    The query then accesses the remaining objects in $g_{p,r}^i$ at $s$ since $h(g_{p,r}^i, r, \rho) = 0$.

    We claim that $v_{p_m} = v_{p_j}$.  Suppose otherwise, then there exists $v_{p_{m-1}} \in g_{p,r}^i$.  We previously determined that $\rho(r,v_{p_m}) \neq \rho(r',v_{p_m})$.    Since the access location for $v_{p_m}$ changes under $r'$, then $r'$ must have added a new replica for $v_{p_m}$ compared to $r$ and $\rho(r',v_{p_m})$ must route the query to that replica to access $v_{p_m}$.
    It follows from Eqn.~\ref{eqn:constrained} that the new replica is placed with the predecessor of $v_{p_m}$ in $p$, that is, $\rho(r',v_{p_m}) = \rho(r',v_{p_{m-1}})$.
    Since $r'$ only adds one object replica for $v_{p_m}$, it holds that $\rho(r',v_{p_{m-1}})= \rho(r,v_{p_{m-1}}) $.  We have so far established that $\rho(r,v_{p_m}) \neq \rho(r',v_{p_m})$, $\rho(r',v_{p_m}) = \rho(r',v_{p_{m-1}})$, and $\rho(r',v_{p_{m-1}}) = \rho(r,v_{p_{m-1}})$. This gives $\rho(r,v_{p_m}) \neq \rho(r,v_{p_{m-1}})$, contradicting that $g_{p,r}^i$ is a server-local path; the claim holds.
    
    Now the only case for $ \rho(r',v_{p_j}) \not= \rho(r,v_{p_j})$ is when  the query accesses $v_{p_j}$ locally under $r'$; otherwise, both $r$ and $r'$ would access the same original copy of $v_{p_j}$. As a result, scheme $r'$ saves one distributed traversal compared to scheme $r$, which routes the query to $d(v_{p_j})$. It now takes $i-1$ distributed traversals to reach $v_{p_j}$ under $r'$, so $v_{p_j}$ is in $g^{i-1}_{p, r'}$.
    We argued previously that $m = j$ is the minimum index where the location of the data accesses with $r$ and $r'$ differs, so to increase the latency of $p$, the data accesses under $r'$ must require at least two extra distributed traversals in $p$ after $v_{p_j}$.

    Let $v_{p_x}$ be the first object in $p$ after $v_{p_j}$ that causes a distributed traversal under $r'$, that is, such that $\rho(r',v_{p_x}) \neq \rho(r',v_{p_{x-1}})$.
    By definition, both $v_{p_j}$ and $v_{p_{x-1}}$ are in $g^{i-1}_{p, r'}$ under $r'$. 

    It follows from Eqn.~\ref{eqn:constrained} that $\rho(r',v_{p_x}) = d(v_{p_x})$, so $r'$ routes the query to the original copy of $v_{p_x}$.
    We have shown that $v_{p_{x-1}}$ is in $g^{i-1}_{p, r'}$, so it takes $i-1$ distributed traversals to reach $v_{p_{x-1}}$ under $r'$ and one more distributed traversal to reach $v_{p_x}$, that is, $v_{p_x}$ is in $g^i_{p, r'}$.

    Let $g_{p,r}^z$ be the server-local subpath including $v_{p_x}$ under $r$.
    By definition of $g_{p,r}^z$, reaching $v_{p_x}$ requires $z$ distributed traversals under $r$.
    Since $x > j$ and $v_{p_j}$ is in $g_{p,r}^i$, it holds that $z \geq i$.
    Therefore, reaching $v_{p_x}$ under $r'$ requires the same number of distributed traversals compared to $r$.

    We now show that there are no additional distributed traversals under $r'$ until we reach the last vertex in $g_{p,r}^z$ by relying on the latency-robustness property. 
    Since $r$ is robust for $p$, it follows that for each object $v_{p_y}$ following $v_{p_x}$ in $g_{p,r}^z$, it holds that $d(v_{p_x}) \in r(v_{p_y})$.
    The same holds under $r'$ since $r'$ is also robust for $p$ according to Lemma~\ref{lem:inductive-robust}.
    This means that a local copy of each $v_{p_y}$ is available at $d(v_{p_x})$ so $\rho(r',v_{p_y})=\rho(r',v_{p_x})$.
    Therefore, reaching the remaining objects in $g_{p,r}^z$ does not require distributed traversals.

    %We have shown that matching all objects until the end of $g_{p,r}^o$ does not require additional distributed traversals in $r'$ compared to $r$.
    In summary, so far we first proved that $r'$ does not introduce additional distributed traversals compared to $r$ to access $v_{p_j}$.
    Then, we showed that if $v_{p_x}$ is the first element after $v_{p_j}$ in $p$ that causes a distributed traversal in $r'$, accessing $v_{p_x}$ or any object in its subpath does not cause additional distributed traversals compared to $r$. 
    The same argument can be applied inductively for the next object in $p$ that requires a distributed traversal under $r'$ after $v_{p_x}$, if it exists.
    Therefore, $r'$ does not require extra distributed traversals compared to $r$ so $h(p,r',\rho) \leq h(p,r,\rho) \leq t_Q$, a contradiction. 
\end{proof}

Lemma~\ref{thm:extend} holds for single-object extensions, but it can be easily applied to situations where $r'$ extends $r$ by multiple objects. 
This follows by applying Lemma~\ref{thm:extend} inductively and by Lemma~\ref{lem:inductive-robust}.

In summary, any extension $r'$ of a replication scheme $r$ that is latency-robust for a path $p$ is also latency-robust for $p$.
Therefore if $r$ is latency-feasible for $p$, then $r'$ is too.
This implies Theorem~\ref{thm:correct-update}.

%In the extended version, we also show that the latency robustness conditionis not necessary for \emph{safe workloads}, i.e., when the height of the causal access tree of a query is $n$ and the latency bound of the query is equal to $0$ or $n-1$.

\subsection{Proof of Theorem~\ref{thm:upward}}
If an optimal replication scheme $r$ contains no additional object replicas compared to $d$, the proof follows directly from Def.~\ref{def:upward_replication}.
If not, a replication scheme $r$ is optimal according to  Eqn.~\ref{eqn:problem} if and only if removing any object replica increases the latency of some query.
This implies that each object replica must be accessed in some causal access path.
Formally, for each replicated object $v \in D$ and for each location of the replica $s \in r(v) \setminus \{d(v)\} $, 
there must exist some causal access path $p'$ of a query $Q\in W$ such that $\rho(r,u) = s$ for the parent $u$ of $v$ in $p'$.
%there must be a parent $u$ of $v$ in some causal access tree $C'$ of a query $Q\in W$ such that $\rho(r,u) = s$ .

Let $v$ be a replicated object in $D$ and $s$ be one of the servers where it is replicated.
We now show that any optimal replicated data placement function $r$ is an upward replication scheme, which according to Def.~\ref{def:upward_replication} implies that there exists some causal access path $p$ such that $\rho(r,u) = \rho(r,v) = s$ for the parent $u$ of $v$ in $p$.

Assume by contradiction that there exists no such $u$ for some replicated object $v \in D$.
This means that for all causal access paths $p$ that accesses the replicated object $v$, if $\rho(r,v) = s$, $\rho(r,u) \neq \rho(r,v)$ for the parent $u$ of $v$ in $p$.
According to Eqn.~\ref{eqn:constrained}, if $\rho(r,u) \neq \rho(r,v)$, then $\rho(r,v) = d(v)$, and this holds for all causal access paths $p$ that include $v$.
However, this contradicts our previous finding that $\rho(r,v) = s \neq d(v)$ for some causal access path $p'$.
    
\end{appendices}

%%
%% If your work has an appendix, this is the place to put it.
% \appendix

% \section{Research Methods}

% \subsection{Part One}

% Lorem ipsum dolor sit amet, consectetur adipiscing elit. Morbi
% malesuada, quam in pulvinar varius, metus nunc fermentum urna, id
% sollicitudin purus odio sit amet enim. Aliquam ullamcorper eu ipsum
% vel mollis. Curabitur quis dictum nisl. Phasellus vel semper risus, et
% lacinia dolor. Integer ultricies commodo sem nec semper.

% \subsection{Part Two}

% Etiam commodo feugiat nisl pulvinar pellentesque. Etiam auctor sodales
% ligula, non varius nibh pulvinar semper. Suspendisse nec lectus non
% ipsum convallis congue hendrerit vitae sapien. Donec at laoreet
% eros. Vivamus non purus placerat, scelerisque diam eu, cursus
% ante. Etiam aliquam tortor auctor efficitur mattis.

% \section{Online Resources}

% Nam id fermentum dui. Suspendisse sagittis tortor a nulla mollis, in
% pulvinar ex pretium. Sed interdum orci quis metus euismod, et sagittis
% enim maximus. Vestibulum gravida massa ut felis suscipit
% congue. Quisque mattis elit a risus ultrices commodo venenatis eget
% dui. Etiam sagittis eleifend elementum.

% Nam interdum magna at lectus dignissim, ac dignissim lorem
% rhoncus. Maecenas eu arcu ac neque placerat aliquam. Nunc pulvinar
% massa et mattis lacinia.

\end{document}